\documentclass[prd,preprint,showpacs,amsmath,amssymb]{revtex4}
\usepackage{epsfig}
\usepackage{dcolumn}
\usepackage{bm}
\usepackage{color}
\usepackage{graphicx}
\usepackage{subfigure}
\usepackage{pstricks}
\usepackage{pst-node}
\usepackage{rotating}
\usepackage{times}
\usepackage{overpic}

\newcommand{\ssb}{\Sigma^0\bar{\Sigma}^0}

\newcommand{\pbar}{\bar{p}}

\newcommand{\GG}{\gamma\gamma}
\newcommand{\pp}{\pi^+\pi^-}

\newcommand{\ppb}{p\bar{p}}

\newcommand{\pppr}{\pi^+\pi^-p\bar{p}}

\newcommand{\psp}{\psi^{\prime}}
\newcommand{\jpsi}{J/\psi}
\newcommand{\ar}{\rightarrow}

\newcommand{\llb}{\Lambda\bar{\Lambda}}
\newcommand{\llbpi}{\llb\pi^0}
\newcommand{\llbeta}{\llb\eta}

\newcommand{\lamb}{\bar{\Lambda}}

\newcommand{\bfg}{\begin{figure}}
\newcommand{\efg}{\end{figure}}
\newcommand{\bitm}{\begin{itemize}}
\newcommand{\eitm}{\end{itemize}}
\newcommand{\bnum}{\begin{enumerate}}
\newcommand{\enum}{\end{enumerate}}
\newcommand{\btbl}{\begin{table}}
\newcommand{\etbl}{\end{table}}
\newcommand{\btbu}{\begin{tabular}}
\newcommand{\etbu}{\end{tabular}}
\newcommand{\bcl}{\begin{center}}
\newcommand{\ecl}{\end{center}}
\newcommand{\bbt}{\bibitem}
\newcommand{\beq}{\begin{equation}}
\newcommand{\eeq}{\end{equation}}
\newcommand{\beqr}{\begin{eqnarray}}
\newcommand{\eeqr}{\end{eqnarray}}

\begin{document}
\title{\boldmath Measurements of the branching fractions for $\jpsi$ and $\psp\ar\llbpi$
  and $\llbeta$ }
\author{
M.~Ablikim$^{1}$, M.~N.~Achasov$^{6}$, O.~Albayrak$^{3}$,
D.~J.~Ambrose$^{39}$, F.~F.~An$^{1}$, Q.~An$^{40}$, J.~Z.~Bai$^{1}$,
Y.~Ban$^{26}$, J.~Becker$^{2}$, J.~V.~Bennett$^{16}$
% ...
\\
\vspace{0.2cm}
(BESIII Collaboration)\\
\vspace{0.2cm} {\it
$^{1}$ Institute of High Energy Physics, Beijing 100049, People's Republic of China\\
$^{2}$ Bochum Ruhr-University, D-44780 Bochum, Germany\\
$^{3}$ Carnegie Mellon University, Pittsburgh, Pennsylvania 15213, USA\\
% ...
$^{6}$ G.I. Budker Institute of Nuclear Physics SB RAS (BINP), Novosibirsk 630090, Russia\\
% ...
$^{16}$ Indiana University, Bloomington, Indiana 47405, USA\\
% ...
$^{26}$ Peking University, Beijing 100871, People's Republic of China\\
% ...
$^{39}$ University of Rochester, Rochester, New York 14627, USA\\
$^{40}$ University of Science and Technology of China, Hefei 230026, People's Republic of China\\
% ...
}
}
\vspace{0.4cm}
\begin{abstract}
  We report on a study of the isospin-violating and conserving decays of
  the $\jpsi$ and $\psp$ charmonium state to $\llbpi$ and $\llbeta$, respectively.
  The data are based on 225 million $\jpsi$ and 106 million $\psp$ events that were
  collected with the BESIII detector. The most accurate measurement of the
  branching fraction of the isospin-violating process
  $\jpsi\ar\llbpi$ is obtained, and the isospin-conserving processes
  $\jpsi\ar\llbeta$ and $\psp\ar\llbeta$ are observed for the first
  time. The branching fractions are measured to be ${\cal
    B}(\jpsi\ar\llbpi) =(3.78\pm 0.27_{\rm stat})\pm 0.29_{\rm sys})\times 10^{-5},~{\cal
    B}(\jpsi\ar\llbeta) =(15.7\pm0.79_{\rm stat}\pm1.52_{\rm sys})\times 10^{-5}$ and
  ${\cal B}(\psp\ar\llbeta) =(2.47\pm 0.34_{\rm stat}\pm0.19_{\rm sys})\times
  10^{-5}$. No significant signal events are observed for $\psp\ar\llbpi$
  decay resulting in an upper limit of the branching fraction of ${\cal B}(\psp\ar\llbpi) < 0.29\times
  10^{-5}$ at the 90\% confidence level. The two-body decay of $\jpsi\ar\Sigma(1385)^0\lamb+c.c.$ is searched for, and the upper limit is $B(\jpsi\ar\Sigma(1385)^0\lamb+c.c.)< 0.81\times 10^{-5}$ at the 90\% confidence level.
\end{abstract}
 \maketitle
\section{Introduction}

The charmonium vector meson, $\jpsi$, is usually interpreted as an SU(3) singlet $c\bar c$ bound states with an isospin $I$=0.
Systematic measurements of its decay rates into final states that are isospin violating are of particular interest,
since these results will provide a sensitive probe to study symmetry-breaking effects in a controlled environment.
In this paper, we present a systematic study of isospin-conserving and violating decays of charmonium vector mesons
into baryonic decays accompanied by a light pseudoscalar meson, namely $\jpsi(\psp)\ar \llbeta$ and $\jpsi(\psp)\ar \llbpi$, respectively.

This work is for a large part motivated by a controversial observation that was made in the past while studying the baryonic decay
of the $\jpsi$. Surprisingly, the average branching fraction of the isospin violating decay of $\jpsi\ar\llbpi$ measured by
DM2~\cite{dm2} and by BESI~\cite{bes1} was determined to be ${\cal B}(\jpsi\ar\llbpi)=(2.2\pm0.6)\times 10^{-4}$,
while the isospin conserving decay mode $\jpsi\ar\llbeta$ was not reported by either experiment.
In 2007, the decays of $\jpsi$ and $\psp$ to the final states with a $\llb$ pair plus a neutral pseudoscalar meson were studied using
58 million $\jpsi$ and 14 million $\psp$ events collected with the BESII detector~\cite{xuxp}. The new measurement suggested that the two
previous studies of $\jpsi\ar\llbpi$ may have overlooked the sizable background contribution from $\jpsi\ar\Sigma^{0}\pi^{0}\lamb+c.c.$. The BESII experiment
removed this type of background contribution and only a few statistically insignificant $\jpsi\ar\llbpi$ signal events remained, resulting in
an upper limit of ${\cal B}(\jpsi\ar\llbpi) < 0.64\times 10^{-4}$. Moreover, the isospin conserving decay mode,
$\jpsi\ar\llbeta$, was observed for the first time with a significance of 4.8$\sigma$. However, signal events of the channels $\psp\ar\llbpi$ and $\psp\ar\llbeta$
were not observed by BESII, and resulted in upper limits of ${\cal B}(\psp\ar\llbpi) < 4.9\times 10^{-5}$ and ${\cal B}(\psp\ar\llbeta) < 1.2\times 10^{-4}$.

In 2009, BESIII collected 225 million $\jpsi$~\cite{ jpsinumber} and 106 million $\psp$~\cite{psipnumber} events. These samples provide
a unique opportunity to revisit these isospin conserving and violating decays with improved sensitivity to confirm the previous
observations in $\jpsi$ decays with BESII. The ambition is to investigate as well the same final states in $\psp$ decays with the new record
in statistics, and look for possible anomalies. A measurement of these branching fractions would be a test of the \textquotedblleft $12\%$" rule~\cite{rule}.
The data allow in addition a search for the two-body decays $\jpsi\ar\Sigma(1385)^0\lamb+c.c.$.

\section{Experimental details}
BEPCII is a double-ring $e^+e^-$ collider that has reached a peak
luminosity of about $0.6\times10^{33} ~\rm{cm}^{-2}\rm{s}^{-1}$ at
the center of mass energy of 3.77 GeV. The cylindrical core of the
BESIII detector consists of a helium-based main drift chamber (MDC),
a plastic scintillator time-of-flight system (TOF), and a CsI(Tl)
electromagnetic calorimeter (EMC), which are all enclosed in a
superconducting solenoidal magnet providing a 1.0 T magnetic field.
The solenoid is supported by an octagonal flux-return yoke with
resistive plate counter muon identifier modules interleaved with
steel. The acceptance for charged particles and photons is 93\% over
4$\pi$ stereo angle, and the charged-particle momentum and photon
energy resolutions at 1 GeV are 0.5\% and 2.5\%, respectively. The
detector is described in more detail in~\cite{BESIII}.

The optimization of the event selection criteria and the estimates of physics
background sources are performed through Monte Carlo (MC) simulations. The BESIII
detector is modeled with the \textsc{geant}{\footnotesize 4}
toolkit~\cite{geant4, geant42}. Signal events are generated according
to a uniform phase-space distribution. Inclusive $\jpsi$ and $\psp$
decays are simulated with the \textsc{kkmc} \cite{kkmc}
generator. Known decays are modeled by the \textsc{evtgen}~\cite{evt1}
generator according to the branching fractions provided by the
Particle Data Group (PDG)~\cite{PDG}, and the remaining unknown decay
modes are generated with the \textsc{lundcharm} model~\cite{evt2}.

\section{Event selection}
\begin{figure}[hbtp]
\subfigure{
\begin{minipage}{0.4\textwidth}
\centering
\includegraphics[width=\textwidth]{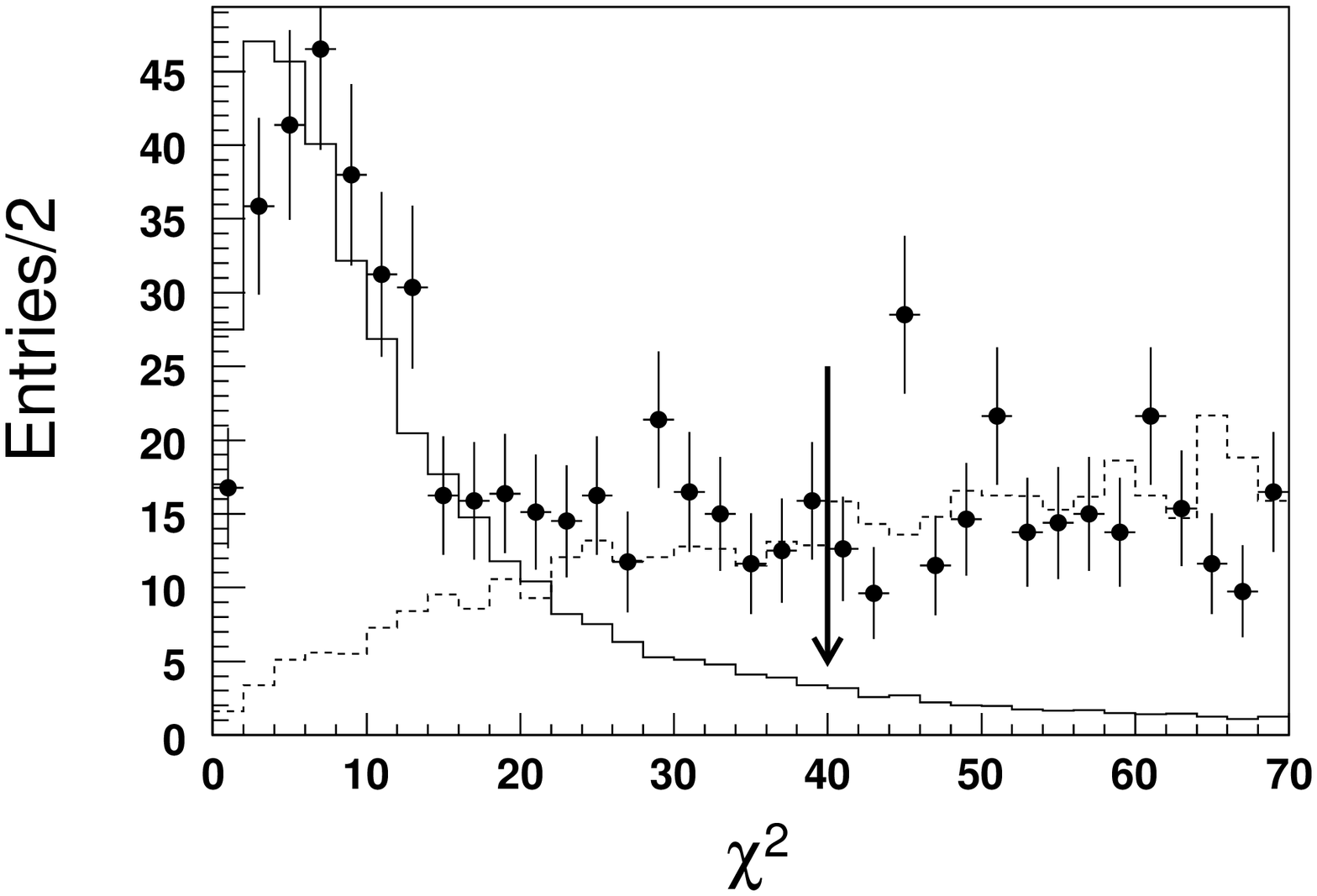}
\put(-30,100){(a)}
\end{minipage}}%
\subfigure{
\begin{minipage}{0.4\textwidth}
\centering
\includegraphics[width=\textwidth]{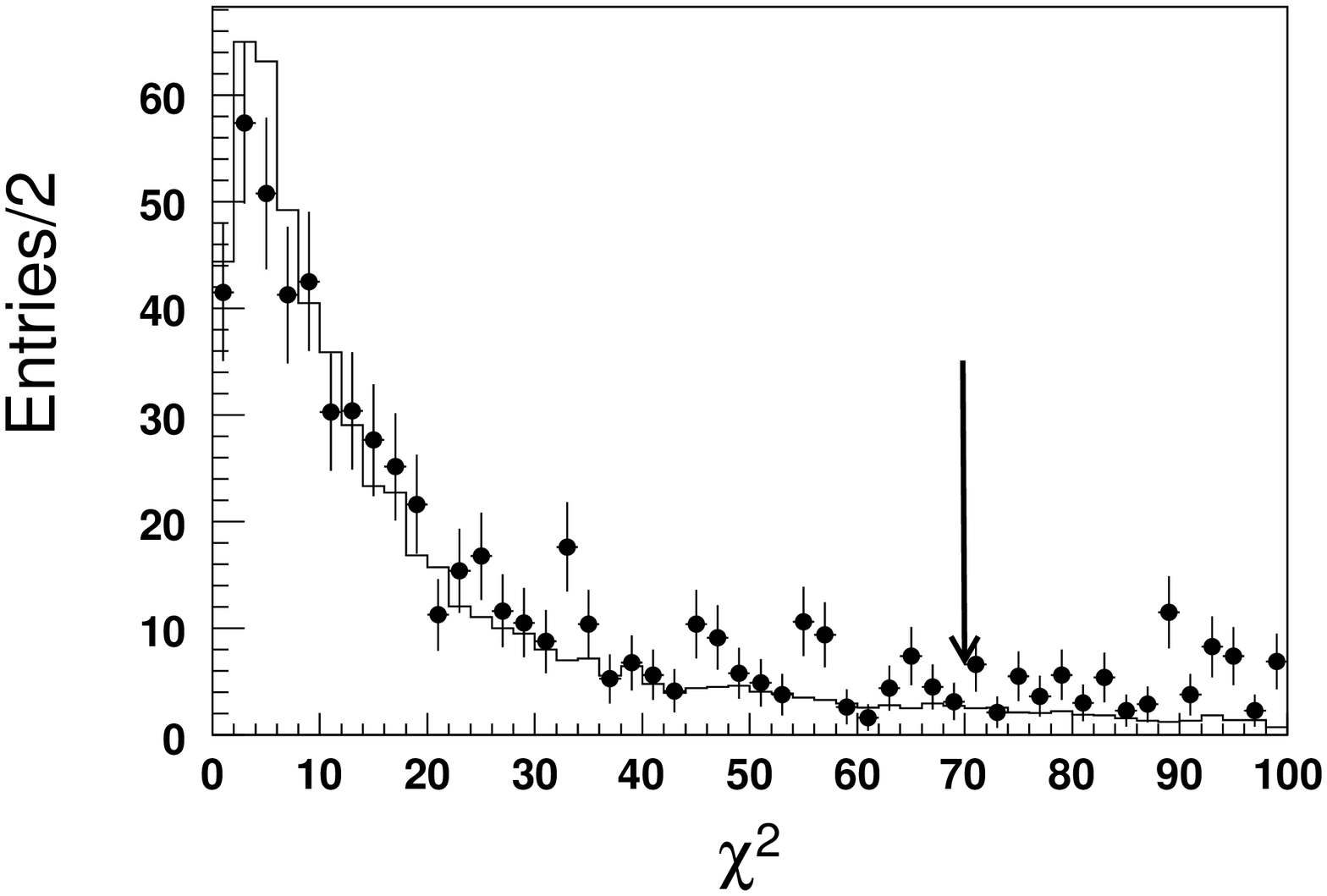}
\put(-30,100){(b)}
\end{minipage}}\\
\subfigure{
\begin{minipage}{0.4\textwidth}
\centering
\includegraphics[width=\textwidth]{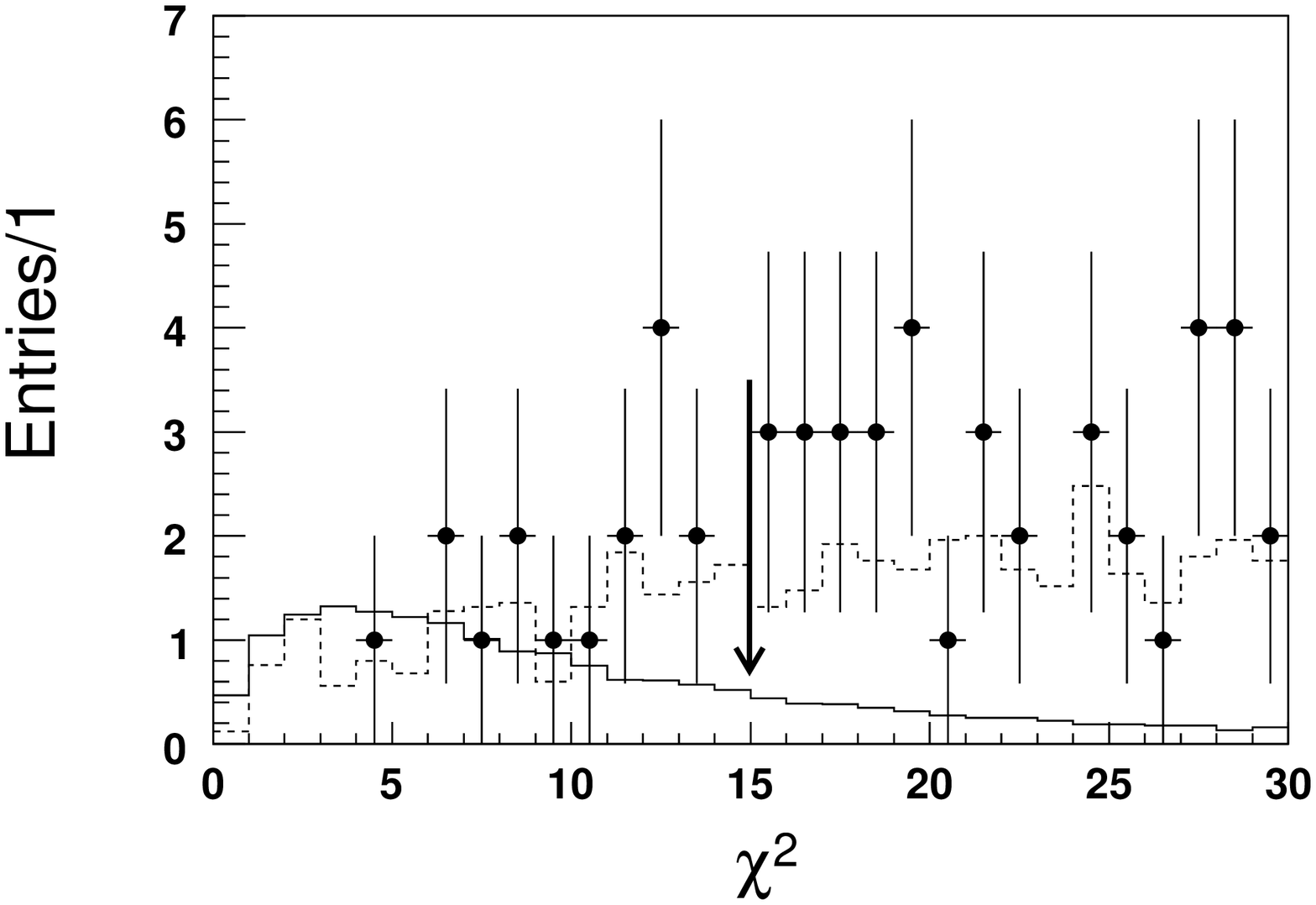}
\put(-30,100){(c)}
\end{minipage}}%
\subfigure{
\begin{minipage}{0.4\textwidth}
\centering
\includegraphics[width=\textwidth]{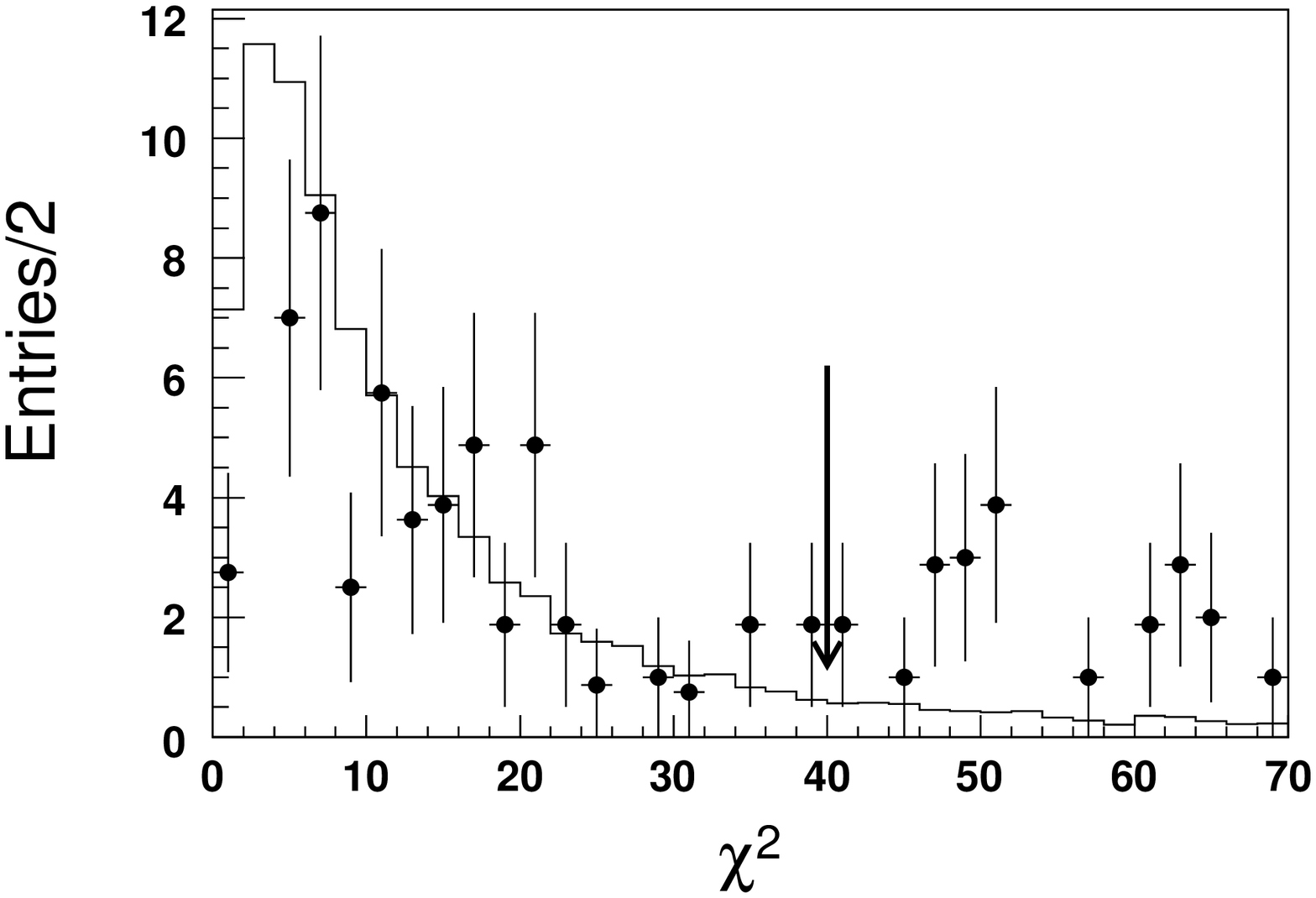}
\put(-30,100){(d)}
\end{minipage}}%
\caption{The $\chi^2_{4C}$ distributions of 4C fits. Dots with error
bars denote data, and the histograms correspond to the result of MC
simulations. (a) $\jpsi\ar\llbpi$. The dashed line is the dominant
background distribution from
$\jpsi\ar\Sigma^0\pi^0\bar{\Lambda}+c.c.$ with MC simulated events,
the arrow denotes the selection of $\chi^2_{4C}<$40. (b)
$\jpsi\ar\llbeta$, the arrow denotes the selection of
$\chi^2_{4C}<$70. (c) $\psp\ar\llbpi$. The dashed line is the
dominant background distribution from
$\psp\ar\Sigma^0\pi^0\bar{\Lambda}+c.c.$ with MC simulated events,
the arrow denotes the selection of $\chi^2_{4C}<$15. (d)
$\psp\ar\llbeta$, and the arrow denotes the selection of
$\chi^2_{4C}<$40.} \label{chi2cut}
\end{figure}

The decay channels investigated in this paper are $\jpsi~(\psp)\ar\llbpi$ and $\jpsi~(\psp)\ar\llbeta$. The final states include
$\Lambda$, $\lamb$ and one neutral pseudoscalar meson ($\pi^0$ or $\eta$), where $\Lambda$~($\lamb$) decays to $\pi^-p$ ($\pi^+\pbar$),
while the $\pi^0$ and $\eta$ decay to $\GG$. Candidate events are required to satisfy the following common selection criteria:
\begin{enumerate}
\item Only events with at least two positively charged and two negatively
  charged tracks are kept. No requirements are made on the impact
  parameters of the charged tracks as the tracks are supposed to
  originate from secondary vertices.
\item The transverse momenta of the proton and anti-proton are
  required to be larger than 0.2~GeV/$c$. Tracks with smaller transverse momenta
  are removed since the MC simulation fails to describe such extremely soft tracks.
\item Photon candidates are identified from the reconstructed showers in the
  EMC. Photon energies are required to be larger than 25~MeV in the
  EMC barrel region ($|\cos\theta|<0.8$) and larger than 50~MeV in
  the EMC end-cap ($0.86<|\cos\theta|<0.92$). The overlapping showers between the
  barrel and end-cap ($0.8<|\cos\theta|<0.86$) are poorly
  reconstructed, therefore, excluded from the analysis. In addition, timing
  requirements are imposed on photon candidates to suppress electronic
  noise and energy deposits from uncorrelated events.
\item The $\Lambda$ and $\lamb$ candidates are identified by
  a reconstruction of decay vertices from pairs of oppositely charged
  tracks $p\pi^{-}$ and $\bar{p}\pi^{+}$~\cite{xum}. At least one
  $p\pi^{-}$ and one $\bar{p}\pi^{+}$ candidate are required to pass the
  $\Lambda$~($\lamb$) vertex fit successfully by looping over all the
  combinations of positive and negative charged tracks. In the case of
  multiple $\Lambda\lamb$ pair candidates, the one with the minimum
  value of
  $(M_{p\pi^{-}}-M_\Lambda)^{2}+(M_{\bar{p}\pi^{+}}-M_{\bar{\Lambda}})^{2}$
  is chosen, where $M_\Lambda$($M_{\bar{\Lambda}}$) is the nominal mass of $\Lambda$($\lamb$), obtained from the PDG~\cite{PDG}.
\item To further reduce the background and to improve the resolution of the reconstructed particle momenta,
  candidate signal events are subjected to a four constraint energy-momentum conservation (4C) kinematic fit under the hypothesis
  of $\jpsi~(\psp)\ar\Lambda\lamb\GG$. In the case of several combinations due to
  additional photons, the one with the best $\chi^2_{4C}$ value is
  chosen.
  In addition, a selection is made on the $\chi^2_{4C}$. Its value is determined by
  optimizing the signal significance $S/\sqrt{S+B}$, where $S~(B)$ is the number of signal (background)
  events in the signal region. This requirement is effective against
  background with one or several additional photons like
  $\jpsi,~\psp\ar\Sigma^0\pi^0\bar{\Lambda}+c.c.~(\Sigma^{0}\ar\gamma\Lambda)$
  or $\jpsi,~\psp\ar\llb+n\gamma~(n\geq 4)$ decays (for instance
  $\jpsi,~\psp\ar\Sigma(1385)^0\bar{\Sigma}(1385)^0$,
  $\Xi^0\bar{\Xi}^0$,~ etc.).
  For $\jpsi\ar\llbpi$, backgrounds are
  suppressed by requiring $\chi^{2}_{4C}<40$ (see
  Fig.~\ref{chi2cut}(a)). For $\jpsi\ar\llbeta$, the requirement is set
  to $\chi^2_{4C}<70$ (see Fig.~\ref{chi2cut}(b)). For $\psp\ar\llbpi$,
  due to the peaking background
  $\psp\ar\Sigma^0\pi^0\bar{\Lambda}+c.c.$ the $\chi^{2}_{4C}$ is
  required to be less than 15 (see Fig.~\ref{chi2cut}(c)). For
  $\psp\ar\llbeta$, we select events with $\chi^2_{4C}<40$ (see
  Fig.~\ref{chi2cut}(d)).
\end{enumerate}

\begin{figure}[hbt]
\subfigure{
\begin{minipage}{0.4\textwidth}
\centering
\includegraphics[width=\textwidth]{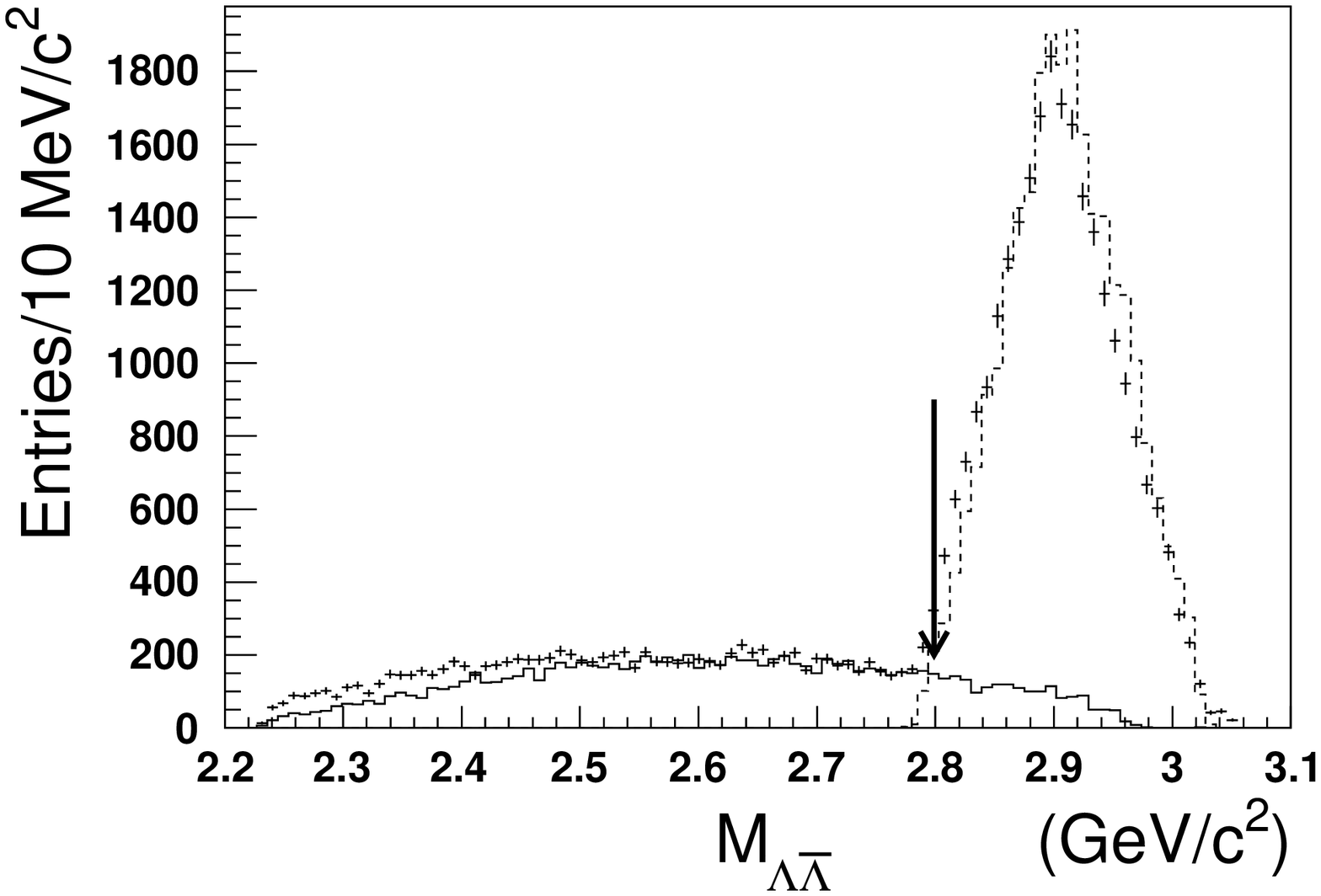}
\put(-140,100){(a)}
\end{minipage}}%
\subfigure{
\begin{minipage}{0.4\textwidth}
\centering
\includegraphics[width=\textwidth]{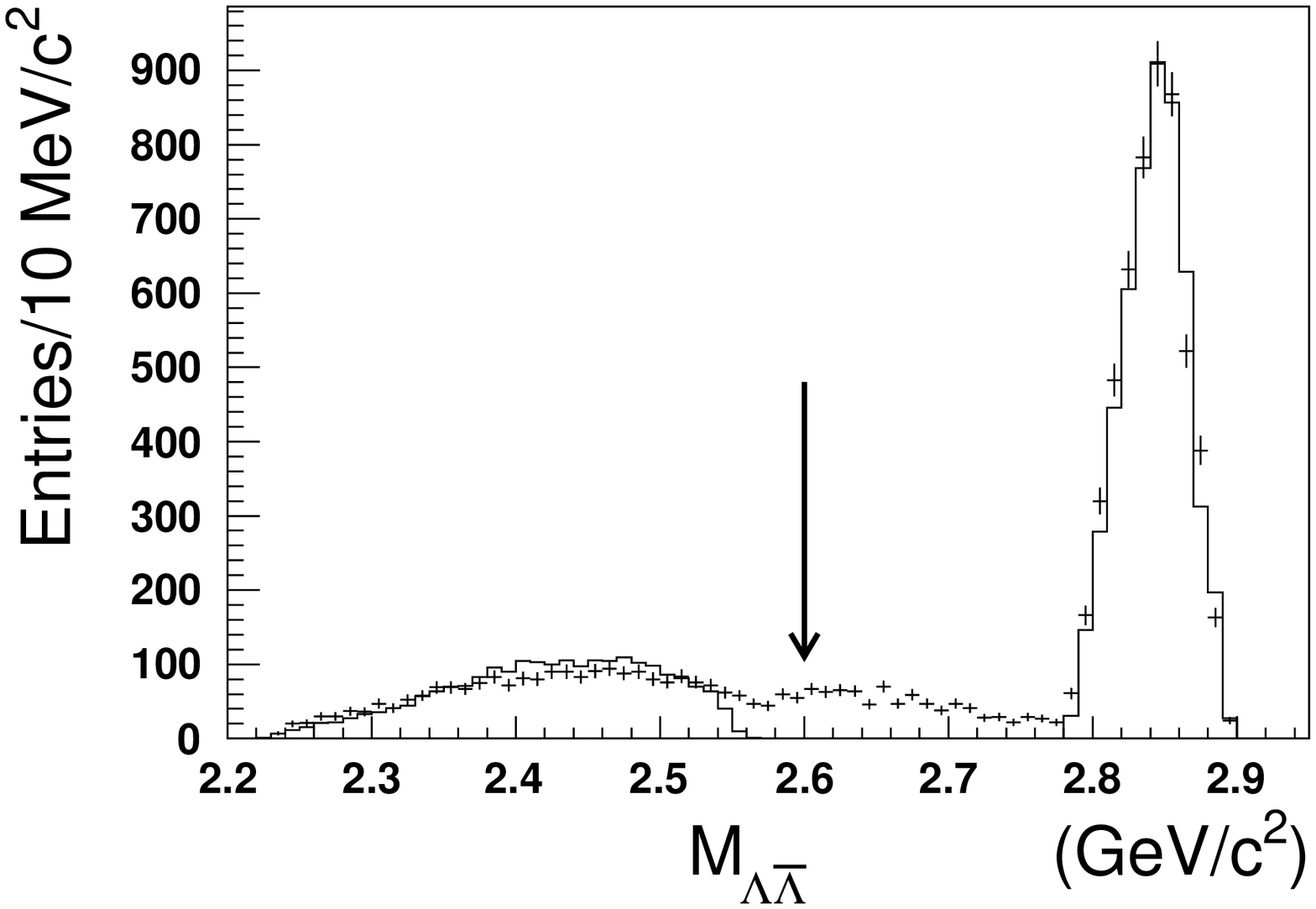}
\put(-140,100){(b)}
\end{minipage}}\\
\caption{The $\llb$ invariant-mass, $M_{\llb}$, distributions for $\jpsi\ar\llb\GG$
  candidates. Dots with errors denote data. The dashed-line shows the
  result of MC simulated events of $\jpsi\ar\ssb$ which is normalized according
  to the branching fraction from the PDG. (a) Histogram shows the MC
  simulated events of $\jpsi\ar\llbpi$, where the arrow denotes the
  selection of $M_{\llb}<$2.8 GeV/$c^2$. (b) Histogram shows the MC
  simulated events of $\jpsi\ar\llbeta$, and the arrow shows the
  selection of $M_{\llb}<$2.6 GeV/$c^2$.}
\label{mllb}
\end{figure}

\begin{figure}[hbt]
\subfigure{ \label{mllbpsip:mini:subfig:a}
\begin{minipage}[b]{0.4\textwidth}
\centering
\includegraphics[width=\textwidth]{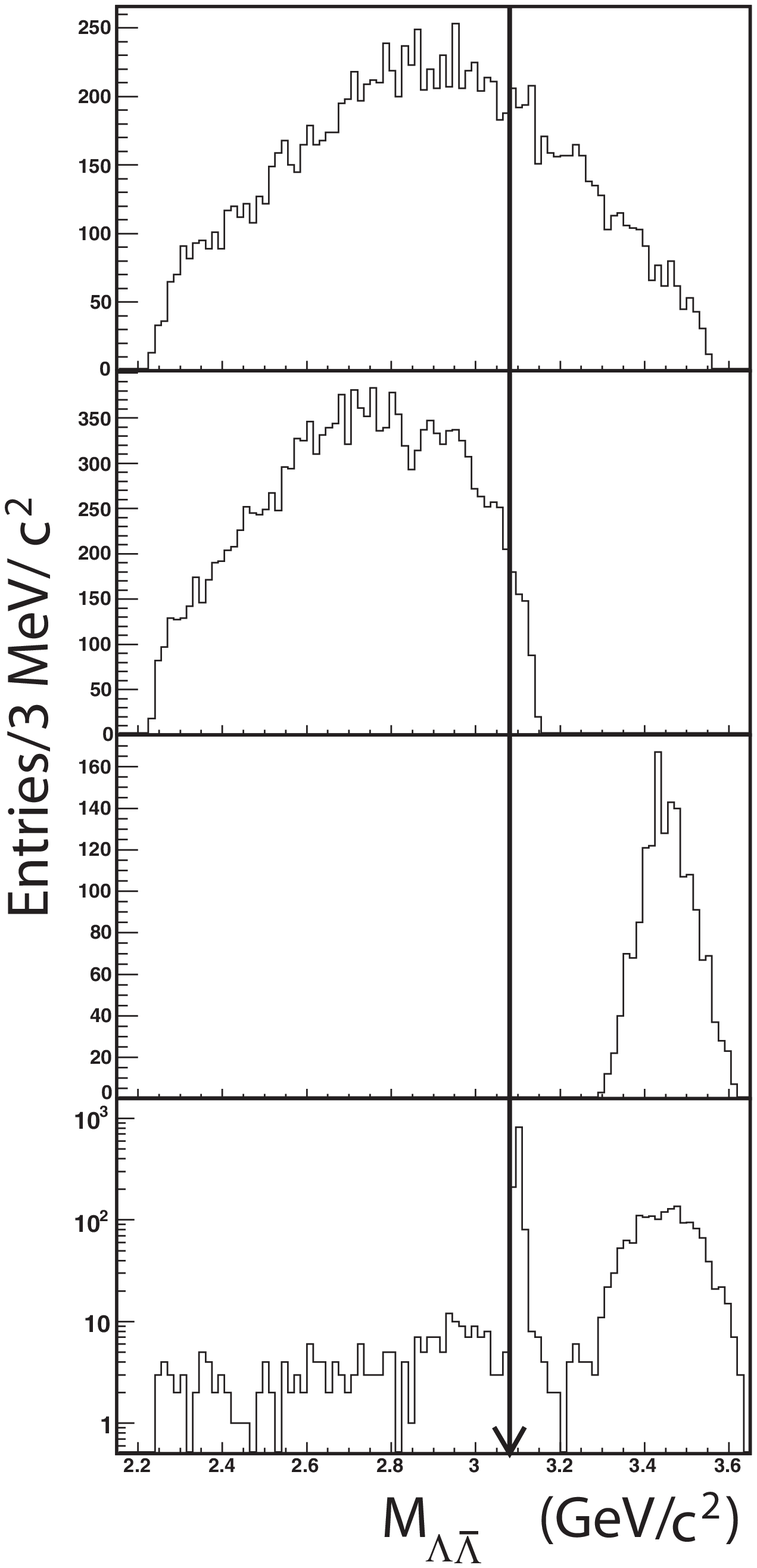}
\put(-150,340){(a)}
\put(-150,260){(b)}
\put(-150,177){(c)}
\put(-150,95){(d)}
\end{minipage}}%
\caption{The $\llb$ invariant-mass, $M_{\llb}$, distributions for $\psp\ar\llb\GG$
  candidates. (a) MC simulated events of $\psp\ar\llbpi$, (b) MC
  simulated events of $\psp\ar\llbeta$, (c) MC simulated events of
  $\psp\ar\ssb$, and (d) data. The arrow denotes the selection of
  $M_{\llb}<$3.08 GeV/$c^2$. The peak around the $\jpsi$ mass is from the
  decay of $\psp\ar\GG\jpsi,\jpsi\ar\llb$.}.
\label{mllbpsip}
\end{figure}

\begin{figure}[hbt]
\subfigure{
\begin{minipage}{0.4\textwidth}
\centering
\includegraphics[width=\textwidth]{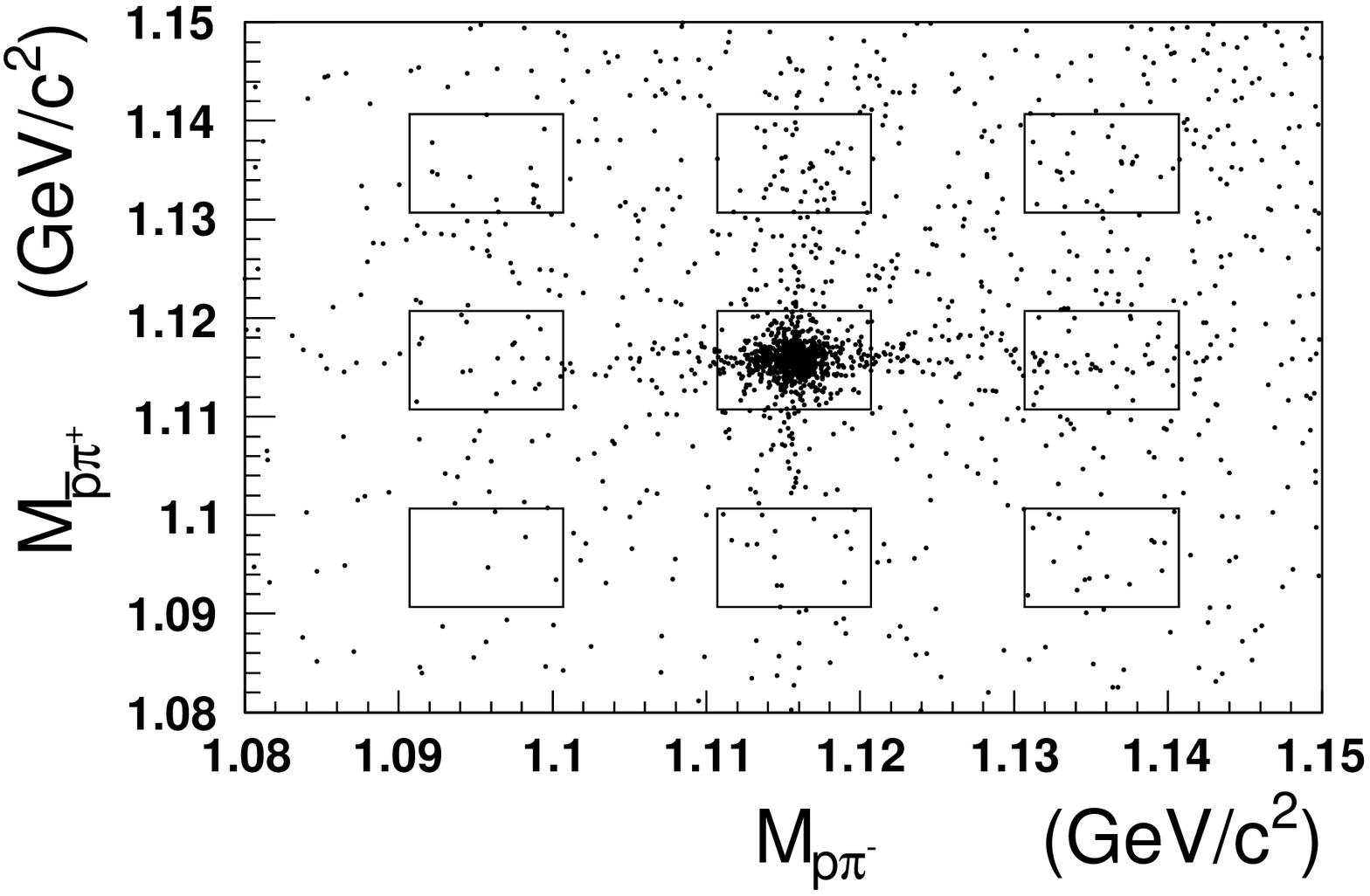}
\put(-155,100){(a)}
\end{minipage}}%
\subfigure{
\begin{minipage}{0.4\textwidth}
\centering
\includegraphics[width=\textwidth]{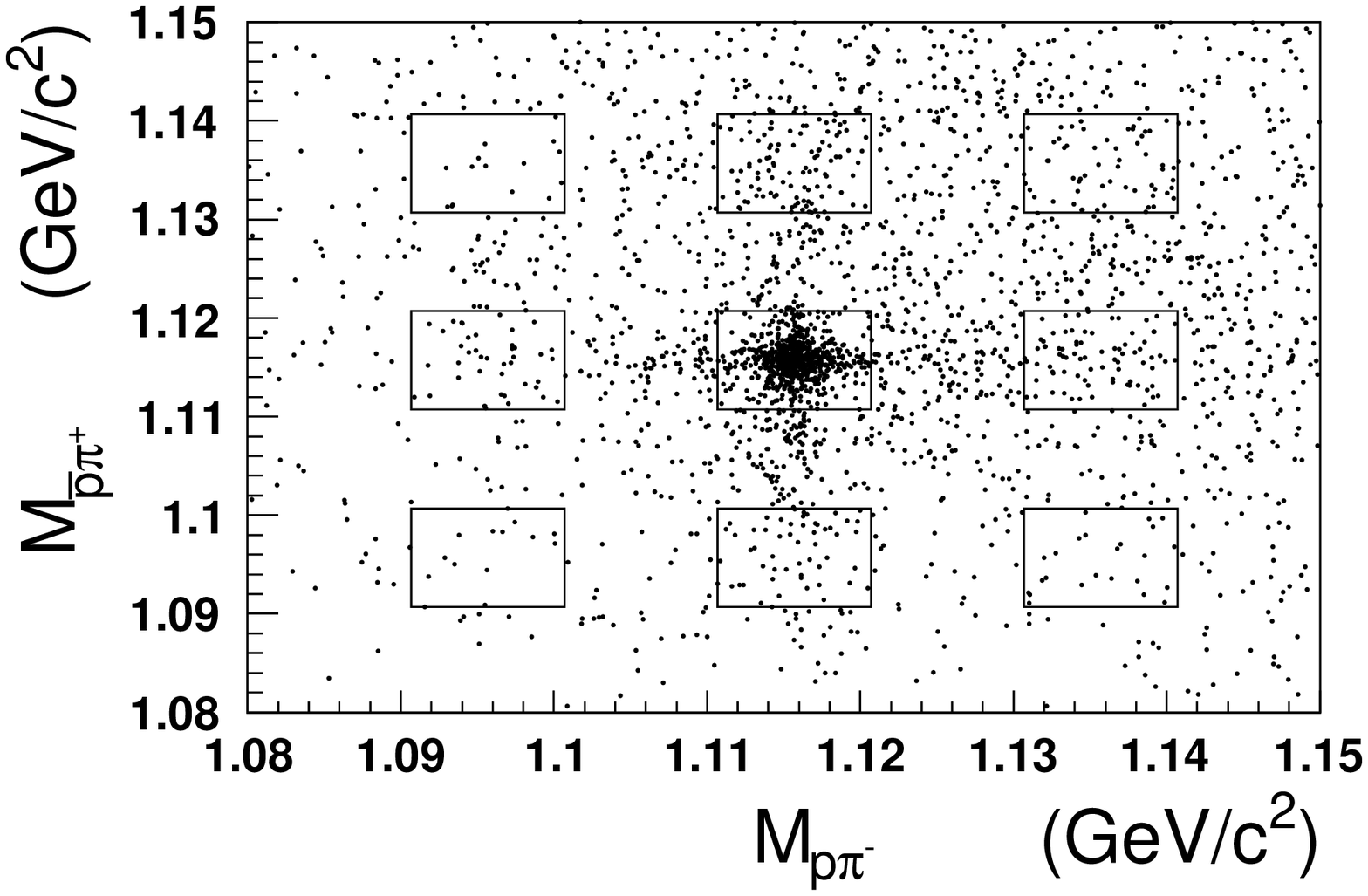}
\put(-155,100){(b)}
\end{minipage}}\\
\subfigure{
\begin{minipage}{0.4\textwidth}
\centering
\includegraphics[width=\textwidth]{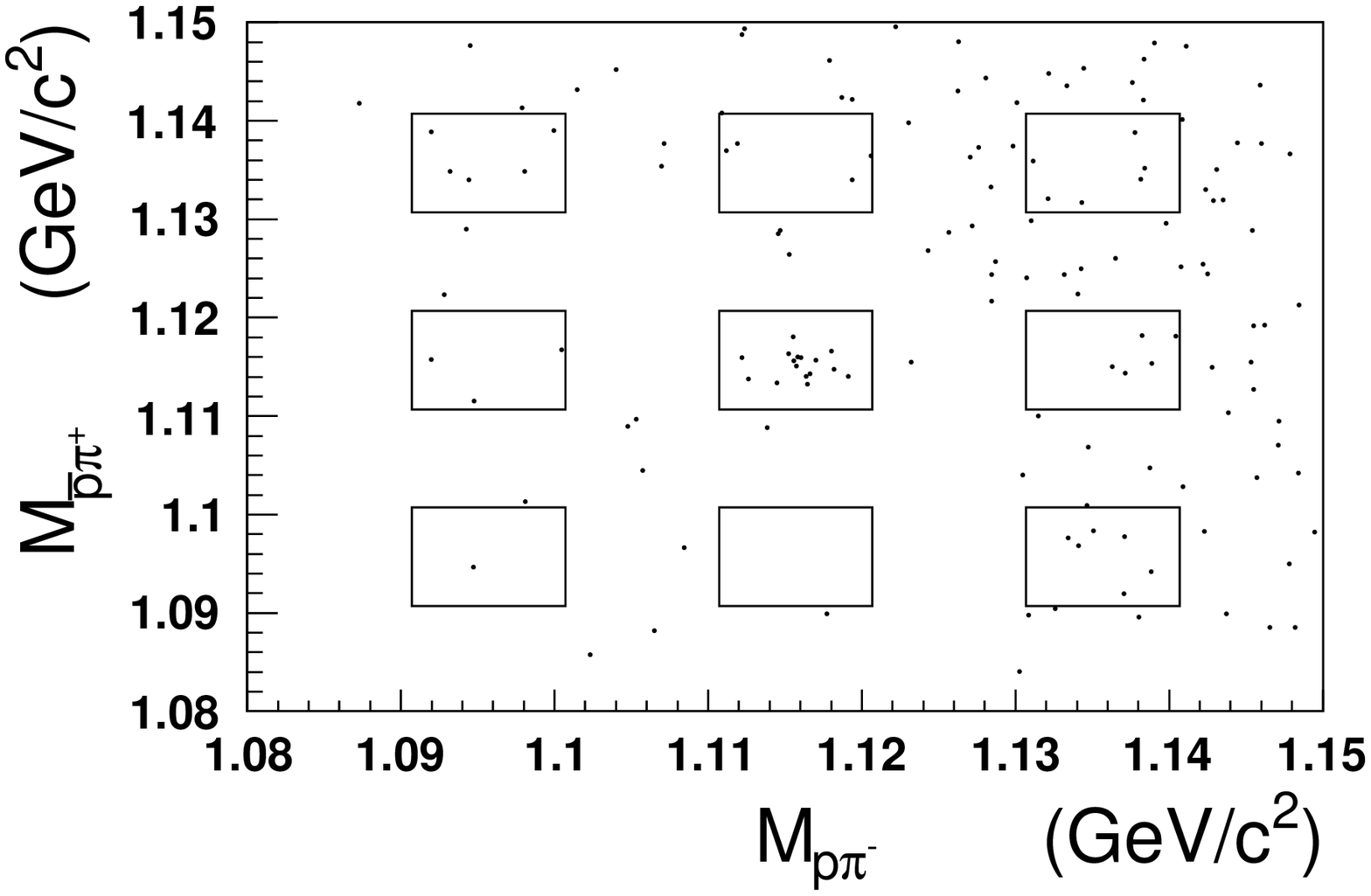}
\put(-155,100){(c)}
\end{minipage}}%
\subfigure{
\begin{minipage}{0.4\textwidth}
\centering
\includegraphics[width=\textwidth]{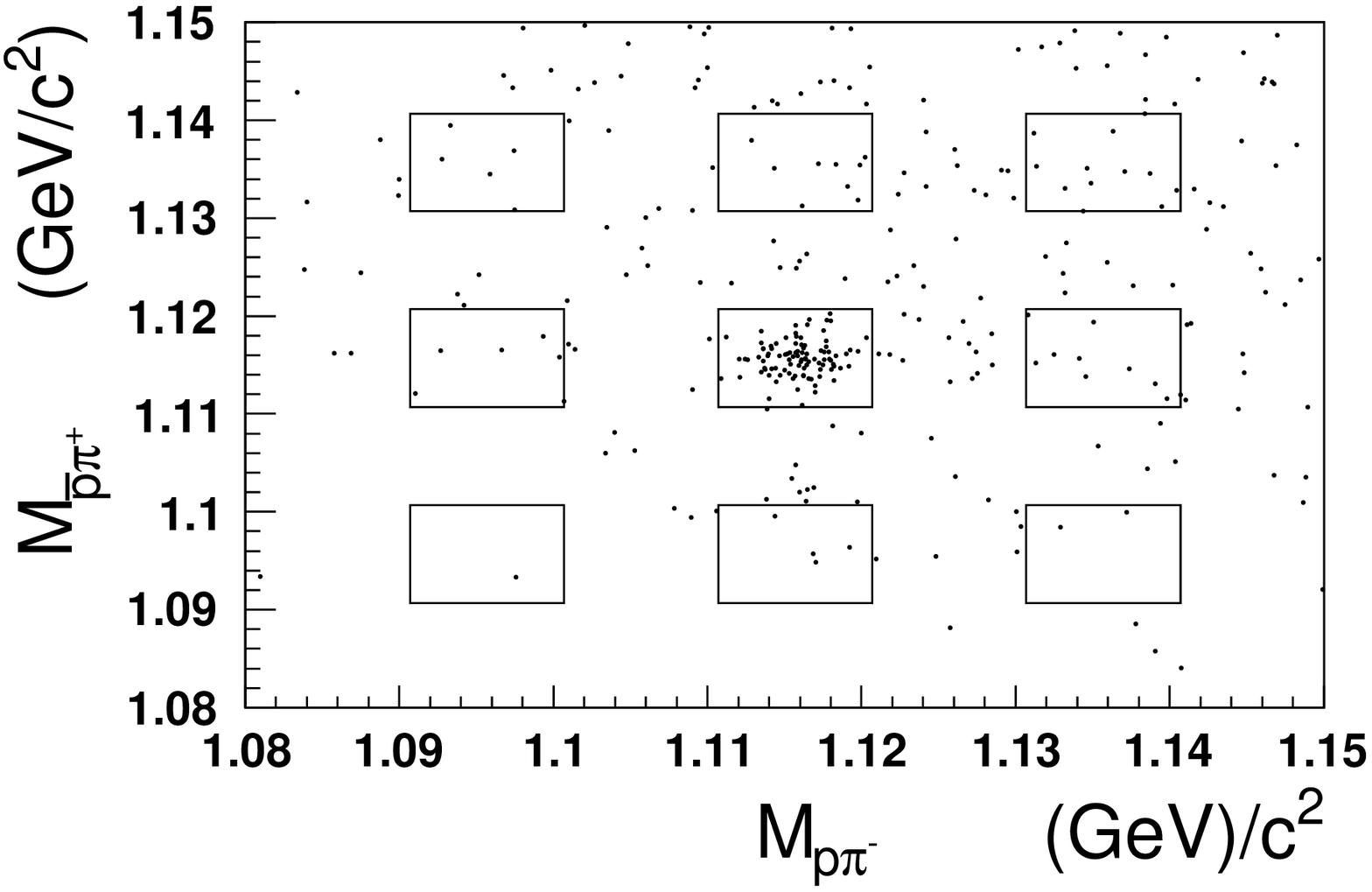}
\put(-155,100){(d)}
\end{minipage}}%
\caption{A scatter plot of $M_{\bar{p}\pi^+}$ versus $M_{p\pi^-}$
  for $\jpsi$ and $\psp$ data. (a) $\jpsi\ar\llbpi$, (b)
  $\jpsi\ar\llbeta$, (c) $\psp\ar\llbpi$, and (d) $\psp\ar\llbeta$.}
\label{scatterplot}
\end{figure}

Followed by the common selection criteria, a further background reduction
is obtained by applying various mass constraints depending on the channel of interest.
To select a clean sample of $\Lambda$ and $\lamb$ signal events, the invariant
masses of $p\pi^{-}$ and $\bar{p}\pi^{+}$ are required to be within
the mass window of $|M_{p\pi}-M_\Lambda|<$~5~MeV/$c^2$. Here, the
invariant mass is reconstructed with improved momenta from the 4C
kinematic fit. The mass resolutions of $\Lambda$ and $\lamb$
are about 1.0 MeV/$c^2$.
For $\jpsi\ar\llbpi$, a mass selection of $|M_{p\pi^0}-1189.0|>$~10~MeV/$c^2$
is used to exclude background from
$\jpsi\ar\Sigma^+\pi^-\bar{\Lambda}+c.c.~(\Sigma^{+}\ar p\pi^{0})$
which can form a peak near the $\pi^{0}$ mass. The background from
$\jpsi\ar\ssb$ is removed by selecting events with $M_{\llb}<2.8$
GeV/$c^2$ as shown in Fig.~\ref{mllb}(a).
For $\jpsi\ar\llbeta$, a selection of events with $M_{\llb}<2.6$
GeV/$c^2$ rejects all background contributions from $\jpsi\ar\ssb$ decays as shown in
Fig.~\ref{mllb}(b).
For $\psp\ar\llbpi$ and $\psp\ar\llbeta$, events must satisfy the condition
$|M_{\pp}^{recoil}-3097|>$~8~MeV/$c^2$ to remove the background from
$\psp\rightarrow\pi^{+}\pi^{-}J/\psi ~(J/\psi\ar p \bar{p}\pi^{0}$ and
$ p \bar{p}\eta)$. The background from $\psp\ar\GG\jpsi
~(\jpsi\ar\llb)$ and $\psp\ar\ssb$ is rejected by the requirement
$M_{\llb}<$ 3.08 GeV/$c^2$. The $\llb$ invariant-mass distributions
for data and MC events from $\psp\ar\llbpi, \llbeta$, and $\ssb$ are
shown in Fig.~\ref{mllbpsip}. The scatter plot of $M_{p\pi^-}$ versus
$M_{\bar{p}\pi^+}$ after applying all selection criteria is shown in
Fig.~\ref{scatterplot}. No visible signal of $\psp\ar\llbpi$ is
observed.

\section{Background study}
Backgrounds that have the same final states as the signal channels
such as $\jpsi,~\psp\ar\ssb,\Sigma^+\pi^-\bar{\Lambda}+c.c.$ are
either suppressed to a negligible level or completely removed.
Background channels that contain one or more photons than the signal channels
like $\jpsi,~\psp\ar\Sigma(1385)^0\bar{\Sigma}(1385)^0,\Xi^0\bar{\Xi}^0$
have very few events passing event selection.
The line shape of the peaking background sources,
$\jpsi,~\psp\ar\Sigma^0\pi^0\bar{\Lambda}+c.c.$, is used in the fitting
procedure to estimate their contributions.
The contribution of remaining backgrounds from non-$\llb$ decays including
$\jpsi,~\psp\ar\pppr\pi^0~(\eta)$ is estimated using sideband studies as
illustrated in Fig.~\ref{scatterplot}. The square with a width of $10$~MeV/$c^2$
around the nominal mass of the $\Lambda$ and $\bar{\Lambda}$
is taken as the signal region.
The eight squares surrounding the
signal region are taken as sideband regions. The area of all the
squares is equal. The sum of events in the sideband squares, $\sum N_{\rm sideband~region}$,
times a normalization factor $f$ is taken as the background contribution in
the signal region. The normalization factor $f$ is defined as \[ f
=\frac{N_{\rm signal~region}}{\sum N_{\rm sideband~region}} .
\]
The normalization factor is obtained from phase-space MC simulations of
$\jpsi~(\psp)\rightarrow p\bar{p}\pi^{+}\pi^{-}\pi^{0}$ or
$\ppb\pp\eta$ with $N_{\rm signal~region}$ as the number of MC events
in the signal region and $\sum N_{\rm sideband~region}$ as the sum of
MC events in the sideband regions.


With 44 pb$^{-1}$ of data collected at a center-of-mass energy of $E_{cm}=3.65$~GeV,
the contribution from the continuum background is determined. From this data sample, no events
survive in the $\pi^{0}$ or $\eta$ mass region in the two-photon invariant-mass, $M_{\GG}$, distribution
after applying all selection criteria. Therefore, we neglect this background.

\section{Signal yields and Dalitz analyses}
\begin{figure}[hbt]
\subfigure{
\begin{minipage}{0.4\textwidth}
\centering
\includegraphics[width=\textwidth]{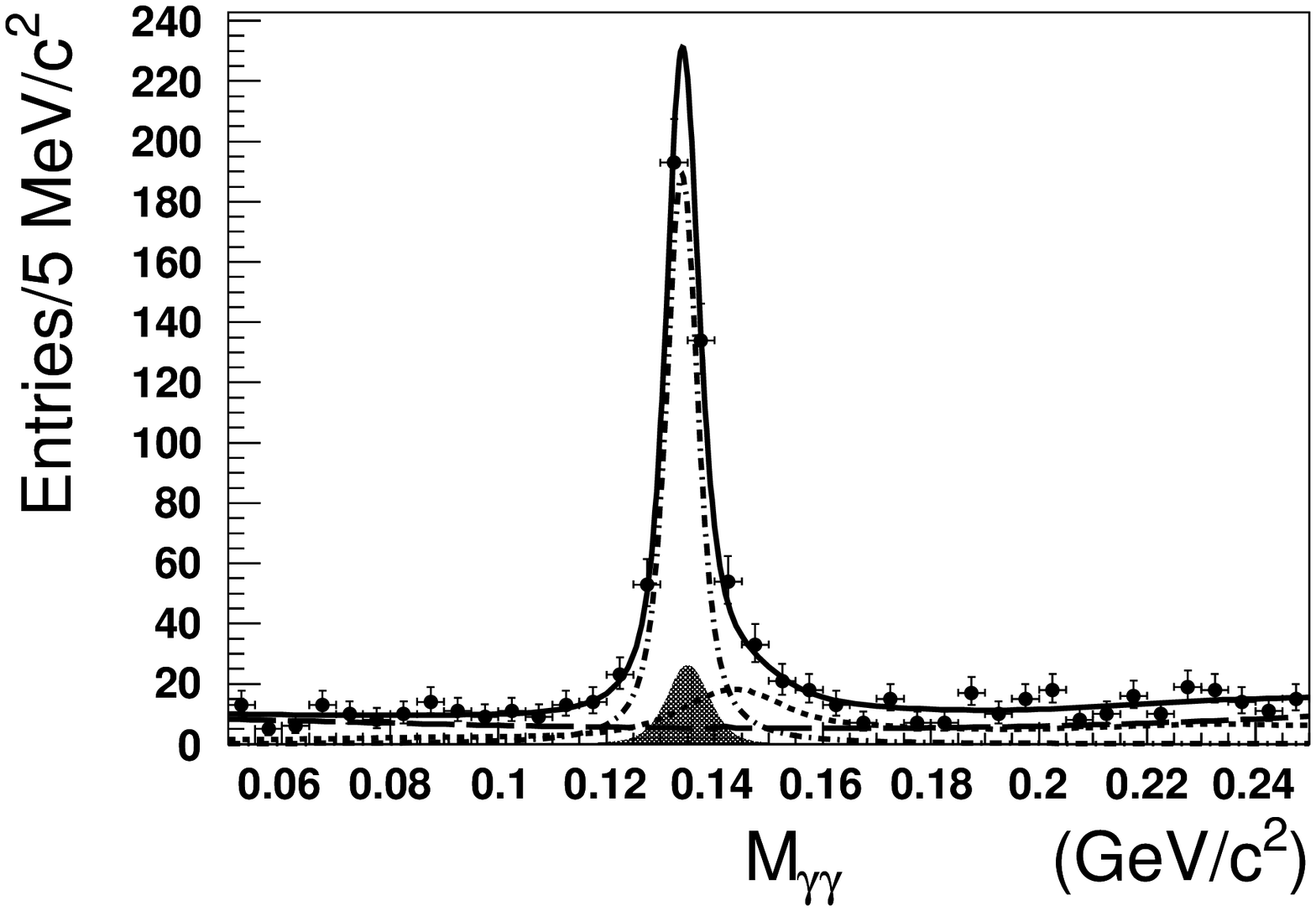}
\put(-145,110){(a)}
\end{minipage}}%
\subfigure{
\begin{minipage}{0.4\textwidth}
\centering
\includegraphics[width=\textwidth]{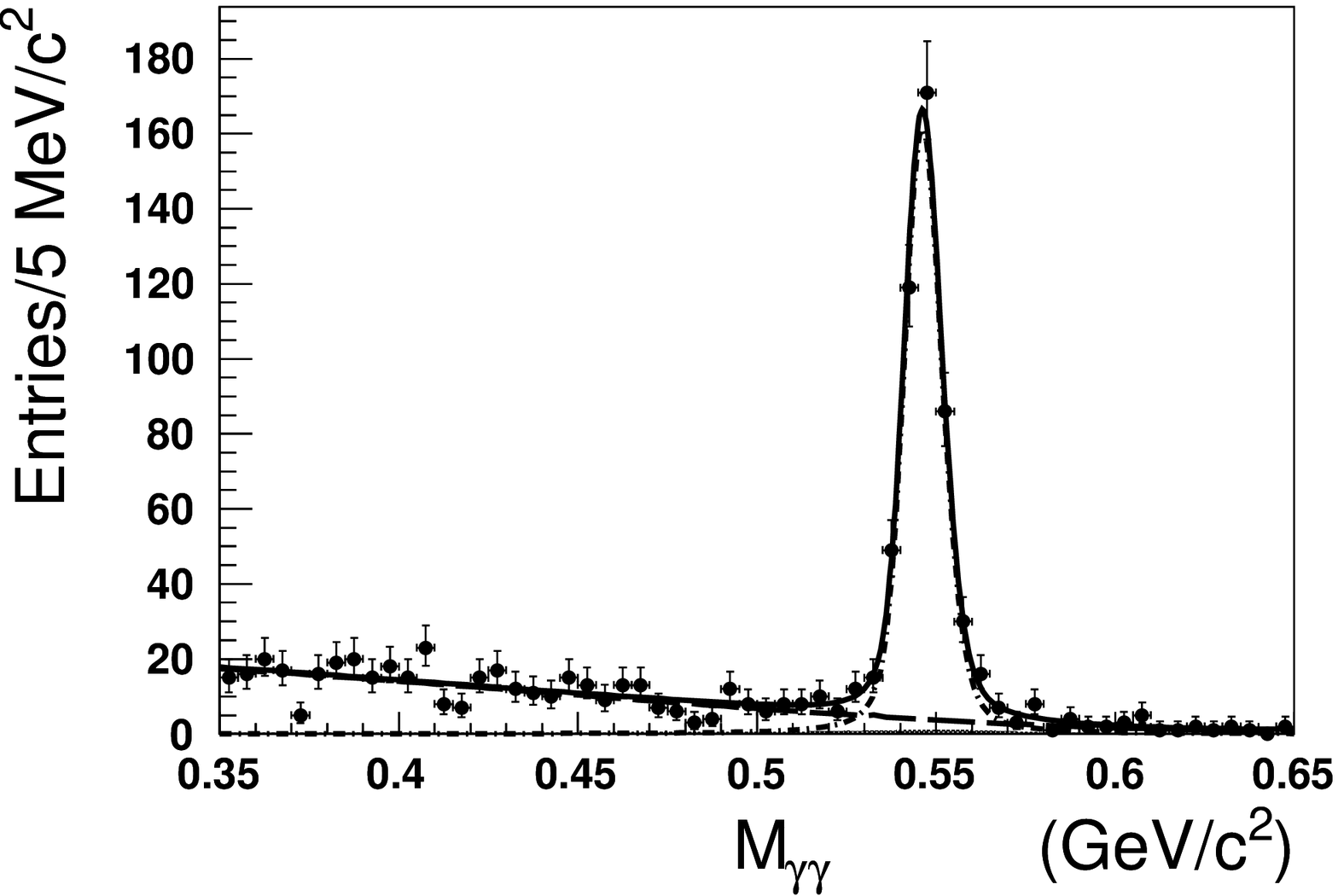}
\put(-145,110){(b)}
\end{minipage}}\\
\subfigure{
\begin{minipage}{0.4\textwidth}
\centering
\includegraphics[width=\textwidth]{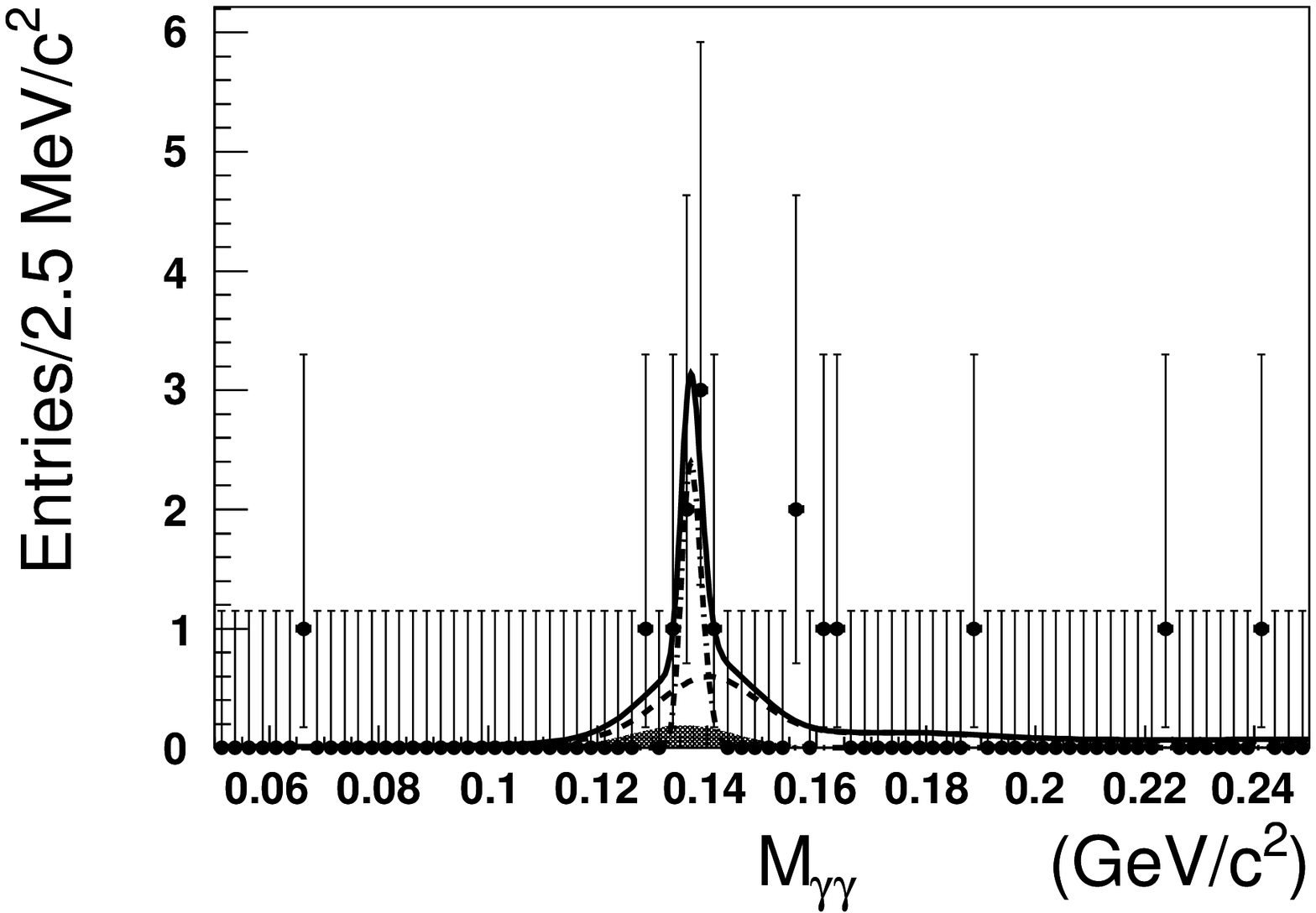}
\put(-145,110){(c)}
\end{minipage}}%
\subfigure{
\begin{minipage}{0.4\textwidth}
\centering
\includegraphics[width=\textwidth]{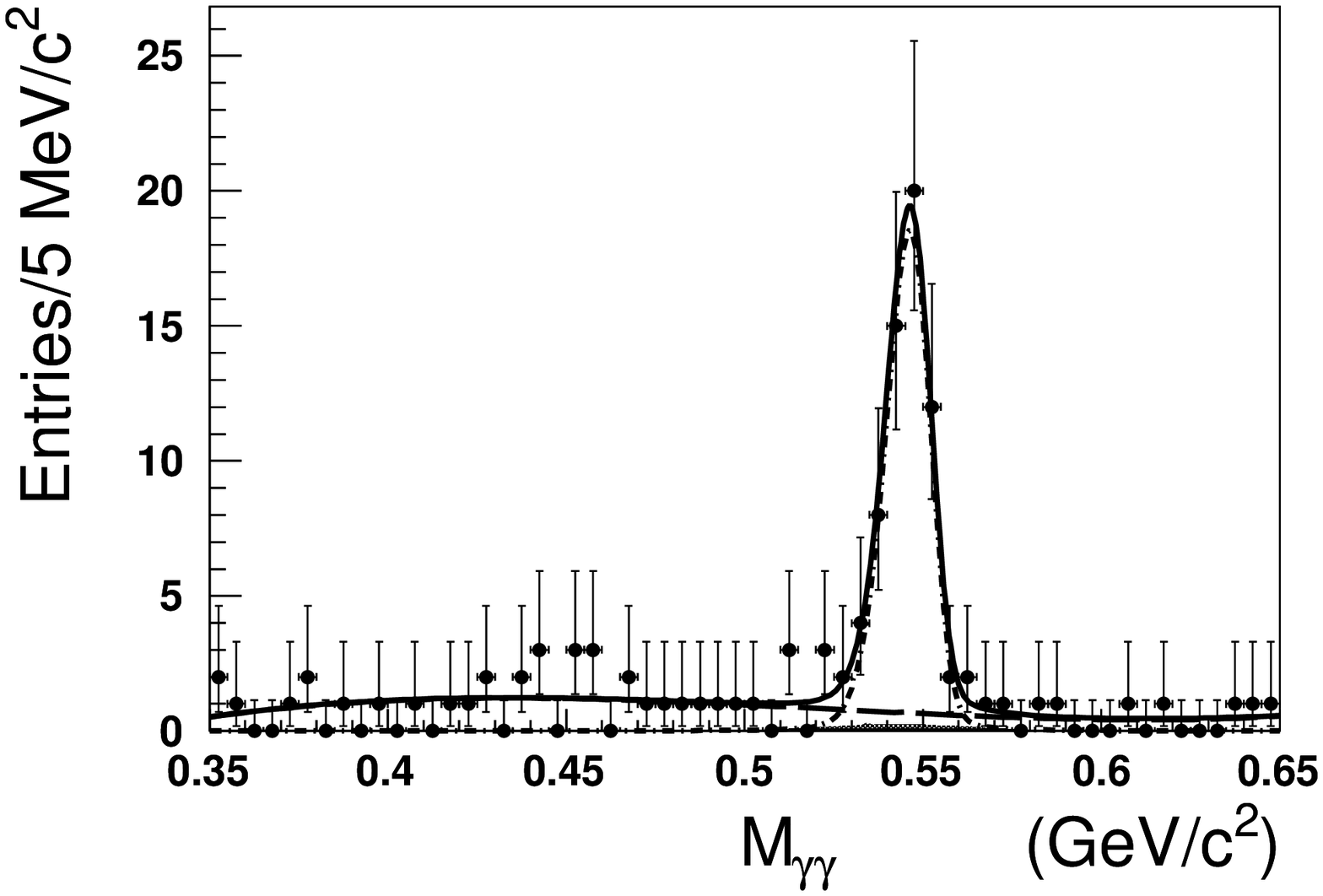}
\put(-145,110){(d)}
\end{minipage}}%
\caption{The two-photon invariant-mass, $M_{\gamma\gamma}$, distributions in the
   $\pi^{0}$ and $\eta$ mass regions for the channels (a) $\jpsi\ar\llbpi$, (b) $\jpsi\ar\llbeta$, (c)
  $\psp\ar\llbpi$, and (d) $\psp\ar\llbeta$. Dots with error bars are
  data. The solid lines are the fit to data, and the dot-dashed lines are the
  signal shape determined from MC simulations. The hatched
  histograms are the background contributions obtained from a normalized sideband analysis.
  The dashed lines in
  $\jpsi,~\psi\ar\llbpi$ correspond to the peaking background from
  $\Sigma^0\pi^0\bar{\Lambda}$. The long dashed lines denote other background contributions
  which are described by Chebychev polynomials.}
\label{fit}
\end{figure}

The $\GG$ invariant-mass spectra of $\jpsi\ar\llbpi,~\llbeta$,
$\psp\ar\llbpi$ and $\llbeta$ of the remaining events after the previously
described signal selection procedure are shown in Fig.~\ref{fit}.
A clear $\pi^0$ and $\eta$ signal can be observed in the $\jpsi$ data.
The $\psp$ data set shows a significant $\eta$ signal, but lacks a
pronounced peak near the $\pi^0$ mass.

The number of signal events are extracted by fitting the $M_{\GG}$
distributions with the parameterized signal shape from MC
simulations. For $\jpsi~(\psp)\ar\llbpi$, the dominant peaking
backgrounds from $\jpsi~(\psp)\ar\Sigma^0\pi^0\bar{\Lambda}+c.c.$ are
estimated by MC simulation.  The fit also accounts for background estimates
from a normalized sideband analysis.
Other background sources are described by a Chebychev polynomial for all channels
except $\psp\ar\llbpi$ where there are too few events surviving.
The fit yields $323\pm 23$ $\pi^0$
events, $454\pm 23$ $\eta$ events in $\jpsi$ data and $60.4\pm 8.4$
$\eta$ events in $\psp$ data. For $\psp\ar\llbpi$, the upper limit on
$N_{\pi^0}$ is 9 at the 90\% confidence level (C.L.) and is determined with a Bayesian
method~\cite{bayes}. For $\psp\ar\llbeta$, the change in log
likelihood value in the fit with and without the signal function is
used to determine the $\eta$ signal significance, which is estimated
to be 10.5$\sigma$.

\begin{figure}[hbt]
\subfigure{
\begin{minipage}{0.4\textwidth}
\centering
\includegraphics[width=\textwidth]{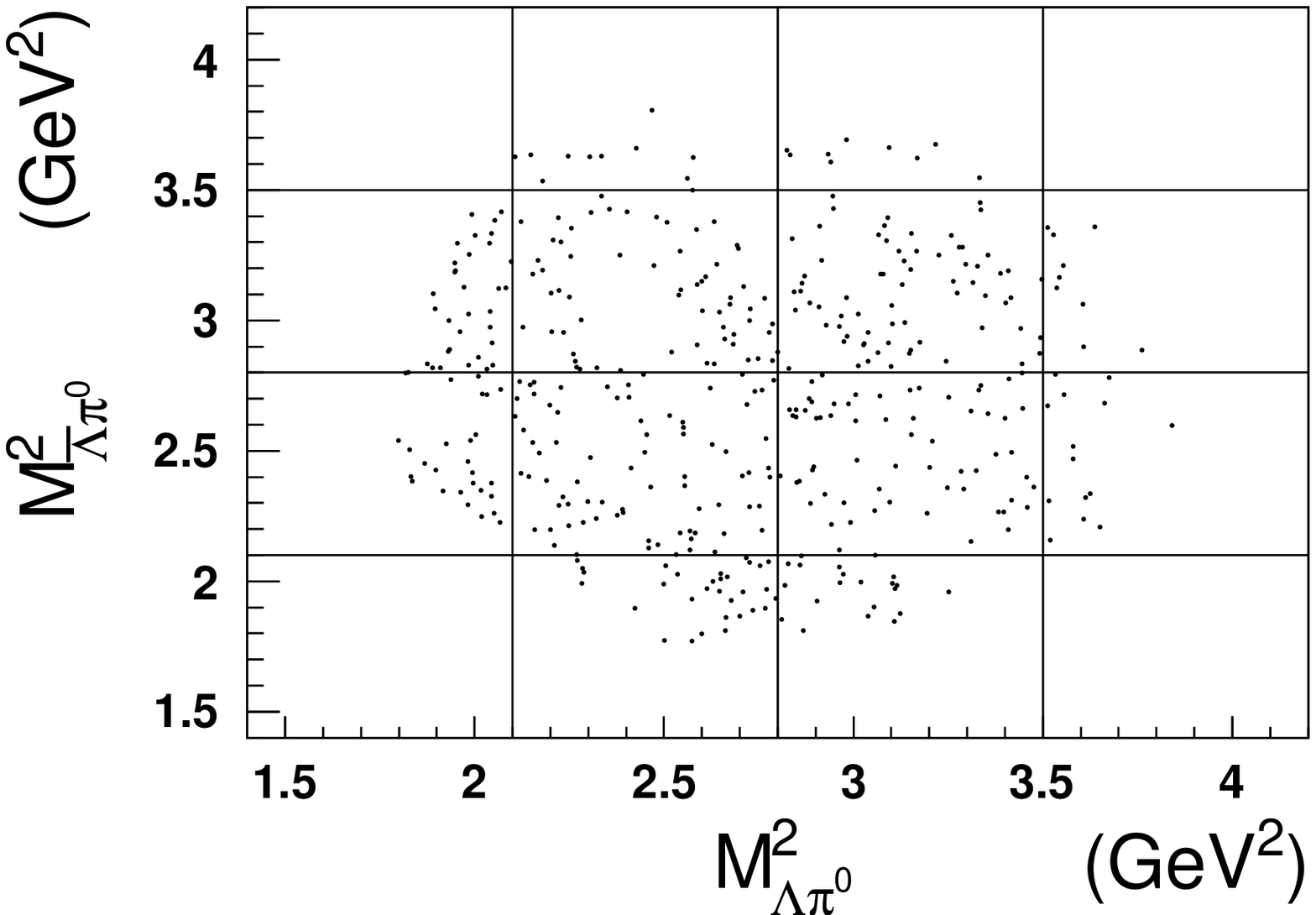}
\put(-145,100){(a)}
\end{minipage}}%
\subfigure{
\begin{minipage}{0.4\textwidth}
\centering
\includegraphics[width=\textwidth]{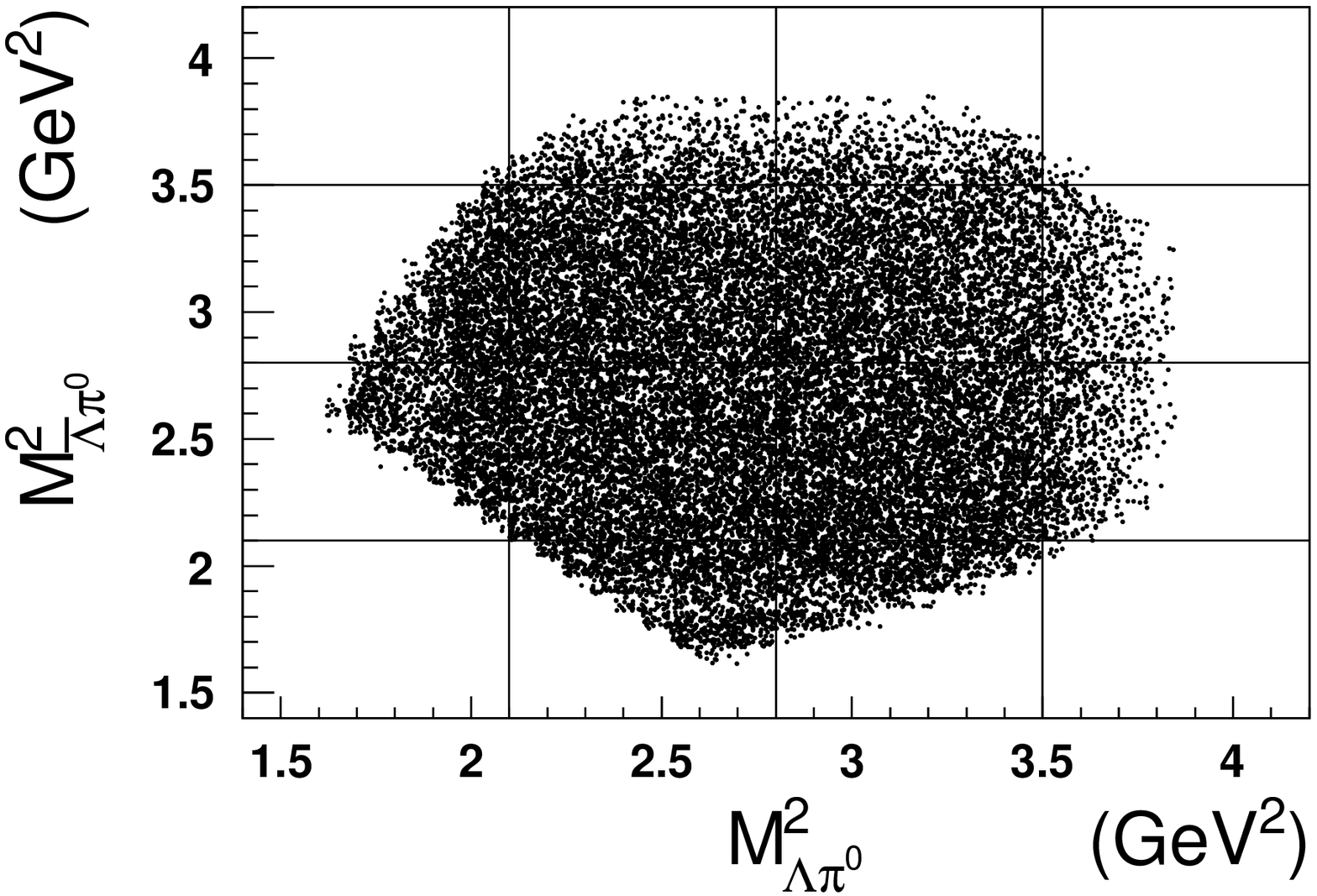}
\put(-145,100){(b)}
\end{minipage}}\\
\subfigure{
\begin{minipage}{0.4\textwidth}
\centering
\includegraphics[width=\textwidth]{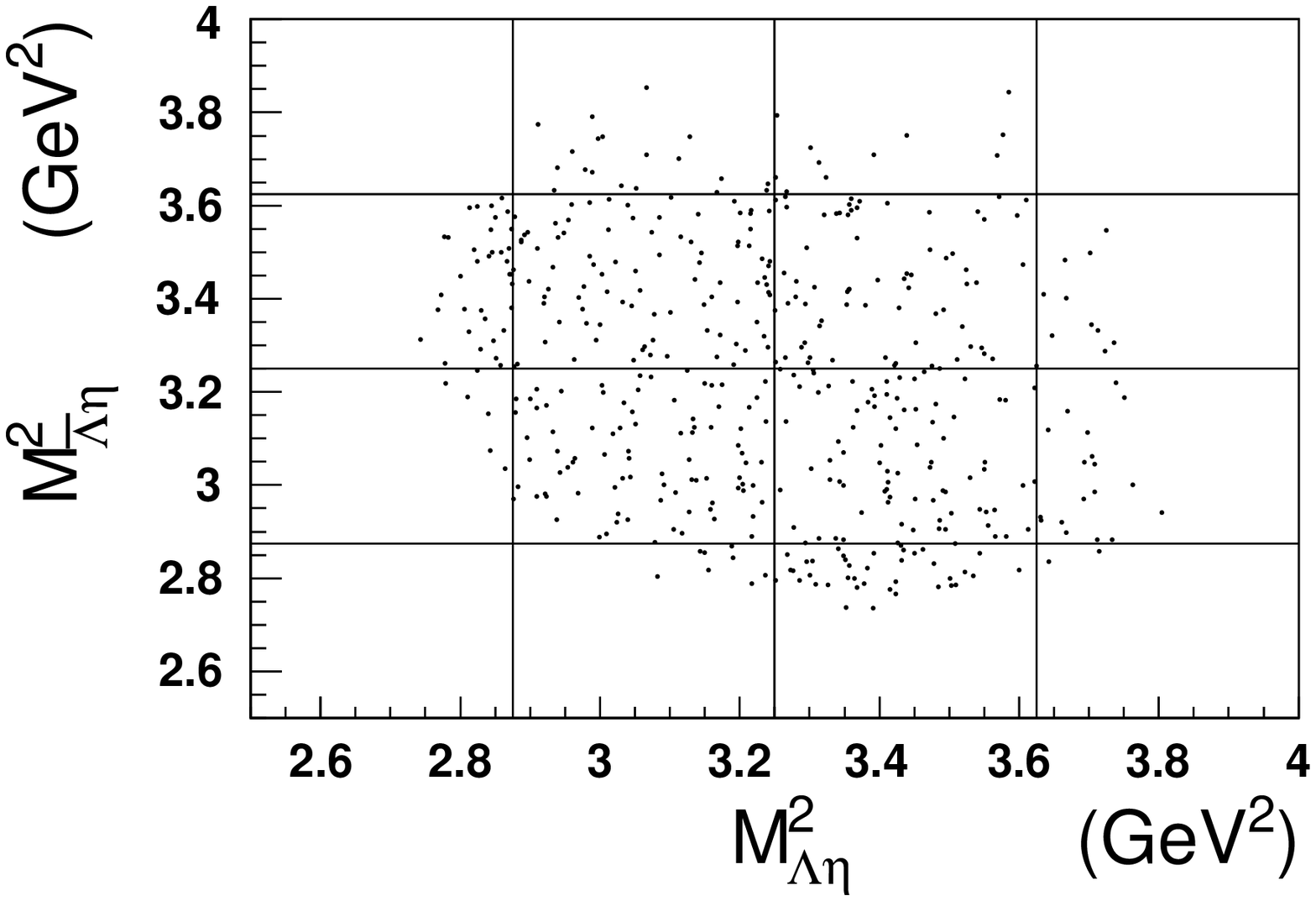}
\put(-145,100){(c)}
\end{minipage}}%
\subfigure{
\begin{minipage}{0.4\textwidth}
\centering
\includegraphics[width=\textwidth]{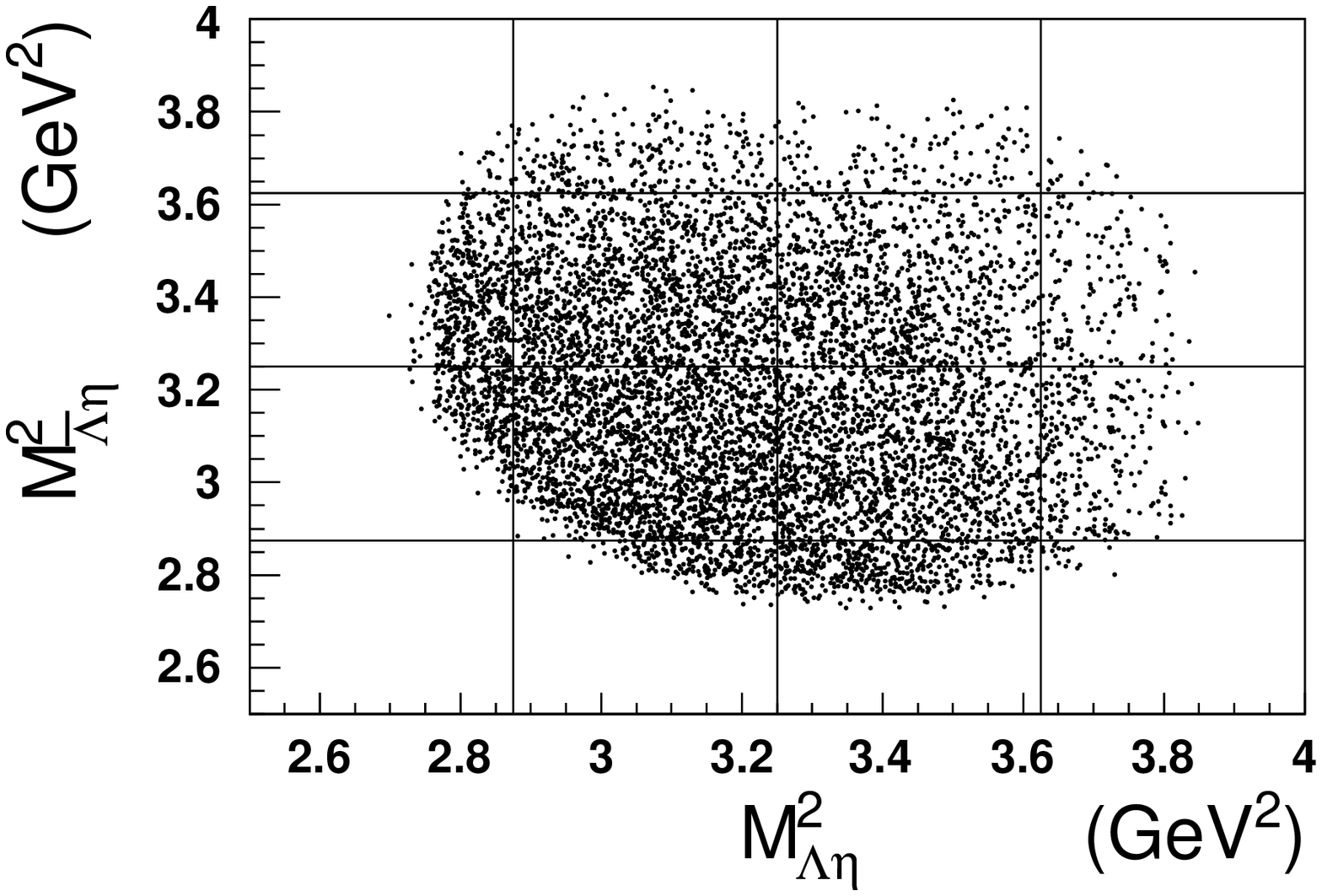}
\put(-145,100){(d)}
\end{minipage}}\\
\subfigure{
\begin{minipage}{0.4\textwidth}
\centering
\includegraphics[width=\textwidth]{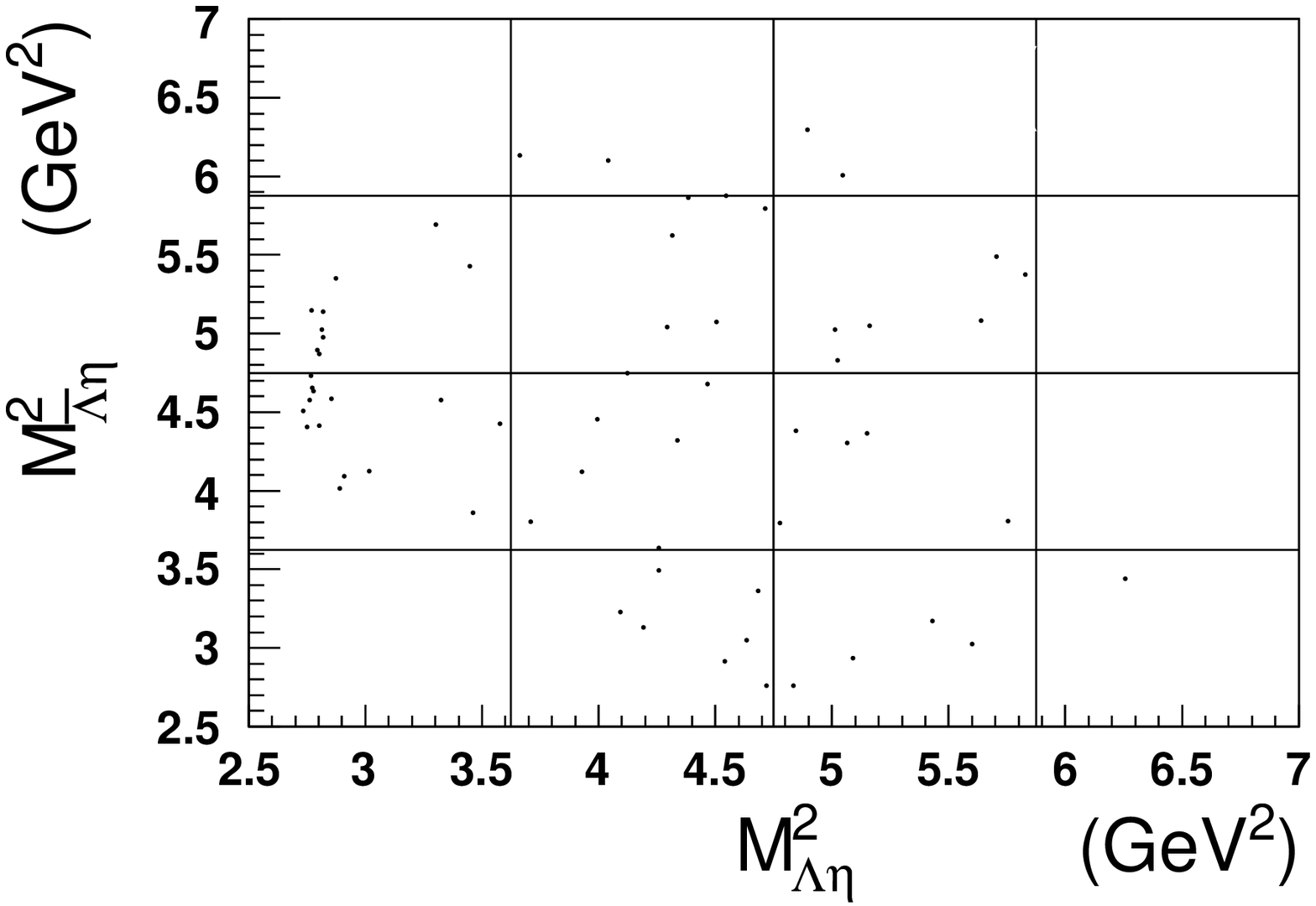}
\put(-145,100){(e)}
\end{minipage}}%
\subfigure{
\begin{minipage}{0.4\textwidth}
\centering
\includegraphics[width=\textwidth]{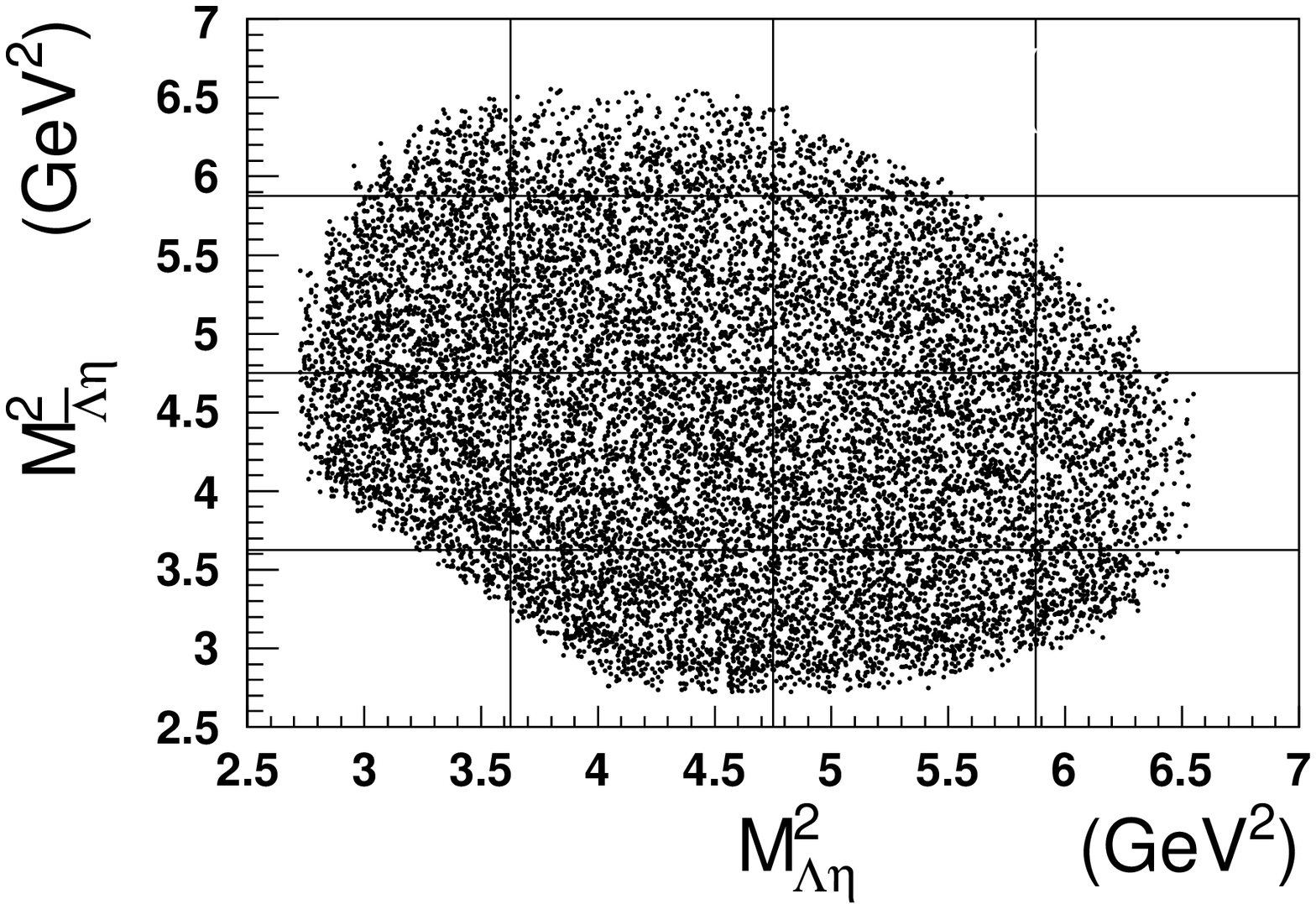}
\put(-145,100){(f)}
\end{minipage}}%
\caption{Dalitz plots of the invariant masses $M_{\bar{\Lambda}\pi^0~(\eta)}^2$ versus
  $M_{\Lambda\pi^0~(\eta)}^2$ for the channels (a) $\jpsi\ar\llbpi$ (data), (b)
  $\jpsi\ar\llbpi$ (MC), (c) $\jpsi\ar\llbeta$ (data), (d)
  $\jpsi\ar\llbeta$ (MC), (e) $\psp\ar\llbeta$ (data), and (f)
  $\psp\ar\llbeta$ (MC). See text for more details.}
\label{fig3}
\end{figure}

To study the existence of intermediate resonance states in the decay of
$\jpsi\ar\llbpi$, $\jpsi\ar\llbeta$ and $\psp\ar\llbeta$ and to validate the
phase-space assumption that was used in the MC simulations, we have performed
a Dalitz plot analysis of the invariant masses involved in the three-body decay.
These results are shown in Fig.~\ref{fig3}. For these plots,
$\pi^0$ and $\eta$ candidates are selected within mass windows of 0.12
GeV/$c^2$$<M_{\GG}<$0.14 GeV/$c^2$ and 0.532 GeV/$c^2$$<M_{\GG}<0.562$
GeV/$c^2$, respectively. In all the Dalitz plots, no clear structures
are observed. A $\chi^2$ test is performed to confirm the consistency
between data and the phase-space distributed MC events. The $\chi^2$ is
determined as follows:
\[
\chi^2=\sum\limits_{i}\frac{(n_i^{data}-n_i^{MC}/g)^2}{n_i^{data}},
\]
where $g$ is the scaling factor between data and MC
$(g=\frac{n_{MC}}{n_{data}})$, $n_i^{data/MC}$ refers to the number of data/MC events
in a particular bin in the Dalitz plot, and the sum runs over all bins.
We divide the Dalitz plots into 8 bins. Boxes with very few
events are combined into an adjacent bin. The $\chi^2/{\rm n.d.f.}$ are equal to 1.1 and
2.1 for $\jpsi\ar\llbpi$ and $\jpsi\ar\llbeta$, respectively, which
validates the usage of a phase-space assumption in the MC simulations.

\begin{figure}[hbt]
\subfigure{ \label{fit1385:mini:subfig:a}
\begin{minipage}[b]{0.4\textwidth}
\centering
\includegraphics[width=\textwidth]{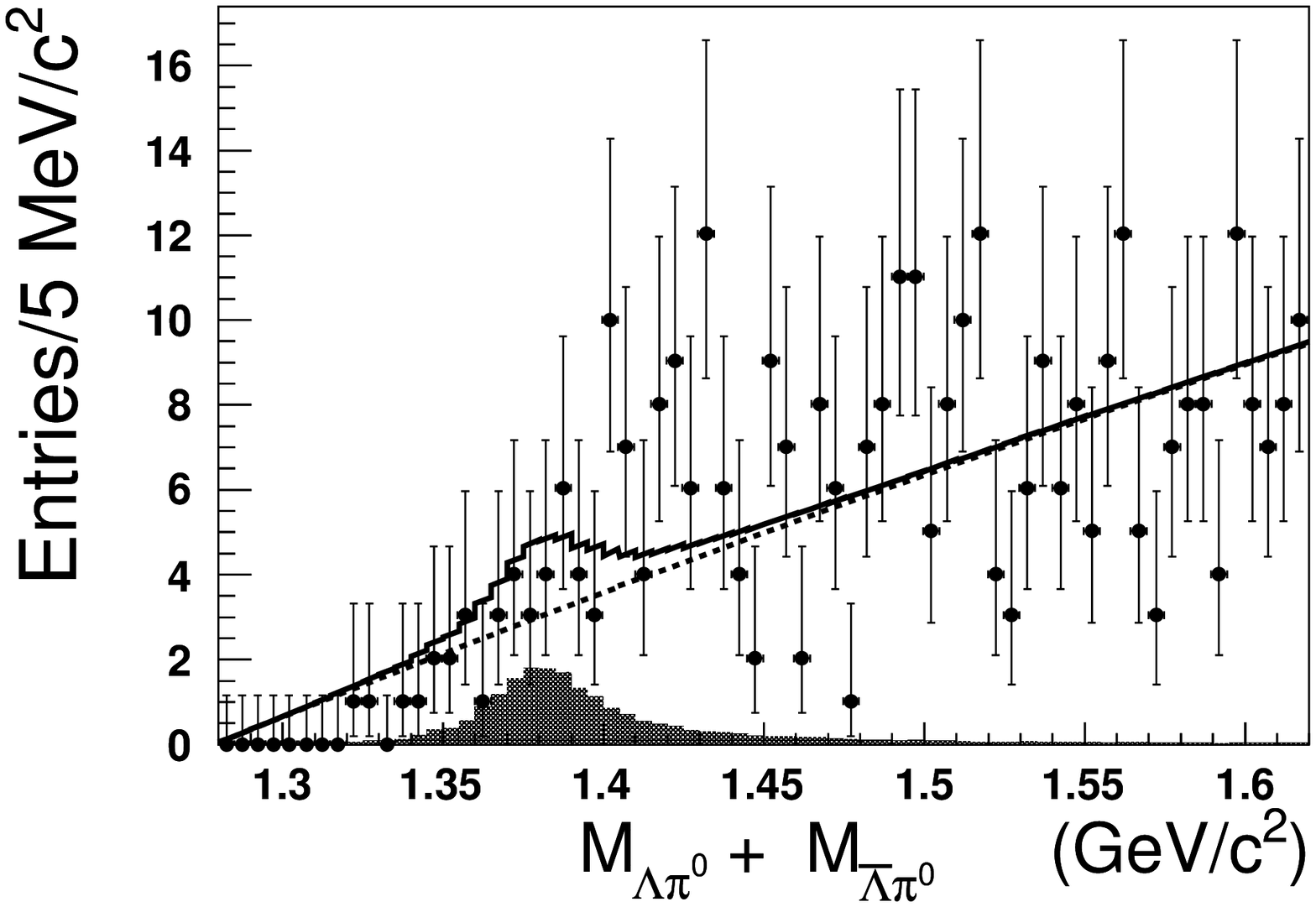}
\end{minipage}}%
\caption{A search for $\Sigma(1385)^0$ events by fitting the combined $M_{\lamb\pi^0}$
  and $M_{\Lambda\pi^0}$ invariant-mass distributions. Dots with error bars are data. The solid
  line is the fit to data. The hatched histogram is the signal
  function obtained from a MC simulation, and the dashed line is the
  background function obtained from a MC simulation of $\jpsi\ar\llbpi$.}
\label{fit1385}
\end{figure}

We have studied the branching fraction of the decay $\jpsi\ar\Sigma(1385)^{0}\bar{\Lambda}+c.c.$
by combining and analyzing the invariant-mass spectra of $\Lambda\pi^0$ and $\lamb\pi^0$ pairs
as depicted in Fig.~\ref{fit1385}. For this analysis, $\pi^0$ events are selected by applying a two-photon
invariant-mass selection of 0.12~GeV/$c^2$$<M_{\GG}<$0.14~GeV/$c^2$.
For the fit, the signal function is taken from a MC simulation of $\jpsi\ar\Sigma(1385)^{0}\bar{\Lambda}+c.c.$, and
the background function is taken from a MC simulation of $\jpsi\ar\llbpi$.
A Bayesian analysis gives an upper limit on the number of
$\Sigma(1385)^{0}\bar{\Lambda}+c.c.$ events of 37 at the 90\% C.L..

\section{Systematic errors}
To estimate the systematic errors in the measured branching fractions of the
channels of interest, we include uncertainties in the efficiency determination of
charged and photon tracks, in the vertex and 4C kinematic fits, in the selection
criteria for the signal and sideband region, and in the fit range. The
uncertainties in the total number of $\jpsi$ and $\psp$ events and in the branching
fractions of intermediate state are considered as well. Below we discuss
briefly the analysis that is used to determine the various sources of
systematic uncertainties.
\bitm
\item Tracking efficiency. We estimate this type of systematic uncertainty by taking the difference between
the tracking efficiency obtained via a control channel from data
with the efficiency obtained from MC simulations. The control sample $\jpsi\rightarrow
pK^{-}\bar{\Lambda} + c.c.$ is employed to study the systematic error
of the tracking efficiency from the $\Lambda~(\lamb)$ decay. For example,
to determine the tracking efficiency of the $\pi^+$ tracks, we select events
with at least three charged tracks, the proton, kaon and anti-proton.
The total number of $\pi^+$ tracks, $N_{\pi^+}^0$, can be determined
by fitting the recoiling mass distribution of the $pK^{-}\bar{p}$
system, $M_{recoil}^{pK^-\bar{p}}$. In addition, one obtains the number
of detected $\pi^+$ tracks, $N_{\pi^+}^1$, by fitting $M_{recoil}^{pK^-\bar{p}}$, after
requiring all four charged tracks be reconstructed. The $\pi^+$
tracking efficiency is simply $\epsilon_{\pi^+} = \frac{N_{\pi^+}^1}{N_{\pi^+}^0}$.
Similarly, we obtained the tracking efficiencies for $\pi^-,~p$, and
$\bar{p}$. With $225\times 10^6$ inclusive MC events, we obtained
the corresponding tracking efficiency for the MC simulation. The tracking
efficiency difference between data and MC simulation is about 1.0\% for each pion
track. This difference is also about 1.0\% for
a proton~(anti-proton) if its transverse momentum, $P_t$, is larger than 0.3~GeV/$c$. The difference
increases to about 10\% for the range $0.2$~GeV/$c$$<$$P_t$$<$$0.3$~GeV/$c$.
Conservatively, we take a systematic error due to tracking of 1\% for each pion. For the
proton~(anti-proton), we use weighted systematic errors, namely 2\% in
$\jpsi\ar\llbpi$, 3.5\% in $\jpsi\ar\llbeta$, 1.5\% in
$\psp\ar\llbpi$, and 2\% in $\psp\ar\llbeta$.

\item Vertex fit. The uncertainties due to the $\Lambda$ and
$\lamb$ vertex fits are determined to be 1.0\% for each
by using the same control samples and a similar procedure as described for the tracking efficiency.

\item Photon efficiency. The photon detection efficiency was studied by comparing the photon efficiency between MC simulation and the control sample
 $\jpsi\ar\rho^0\pi^0$. The relative efficiency difference is about 1\% for each photon~\cite{bianjm}, which value was used as a systematic uncertainty.

\begin{figure}[hbt]
\subfigure{ \label{fig7:mini:subfig:a}
\begin{minipage}[b]{0.4\textwidth}
\centering
\includegraphics[width=\textwidth]{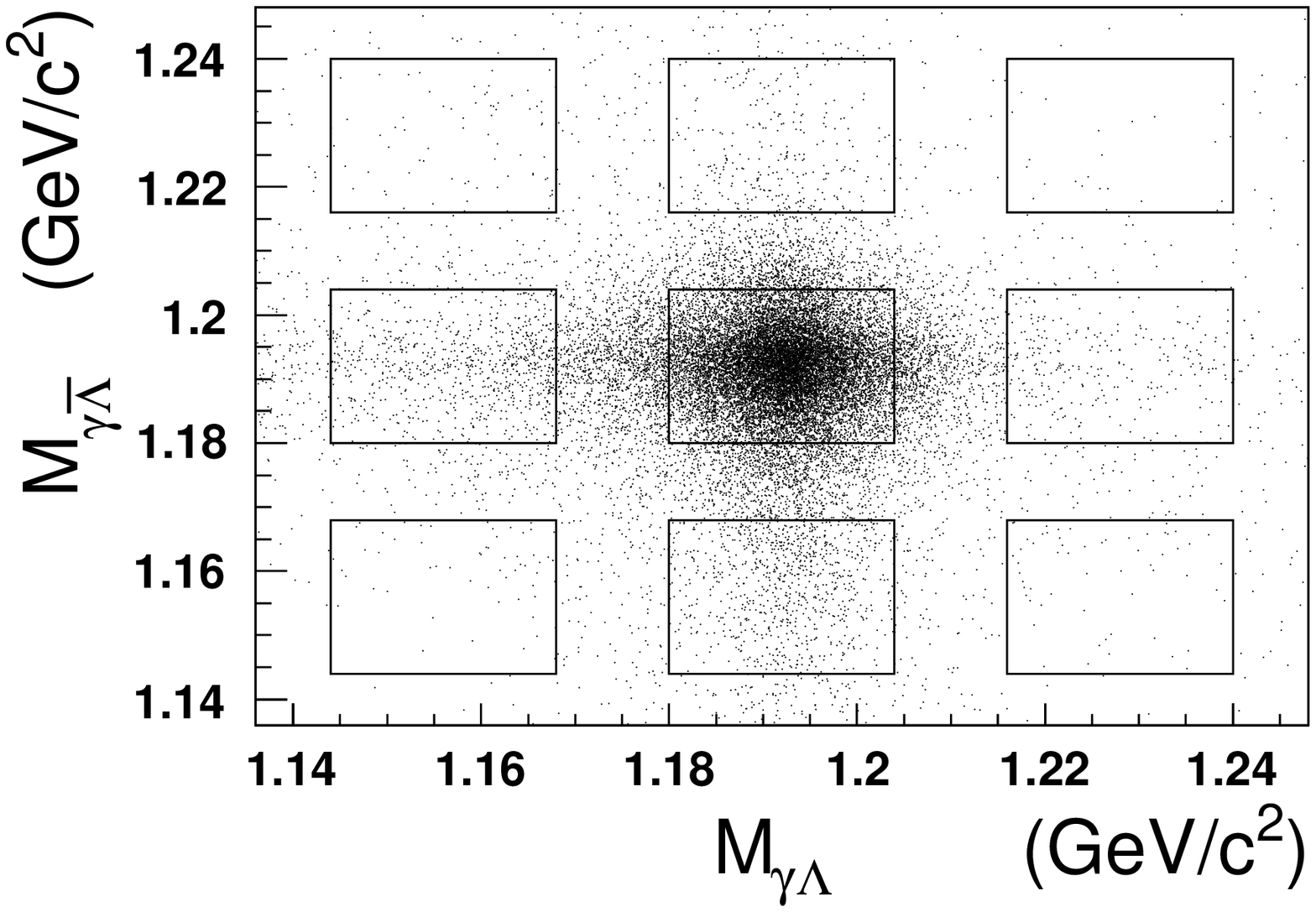}
\end{minipage}}%
\subfigure{ \label{fig7:mini:subfig:b}
\begin{minipage}[b]{0.4\textwidth}
\centering
\includegraphics[width=\textwidth]{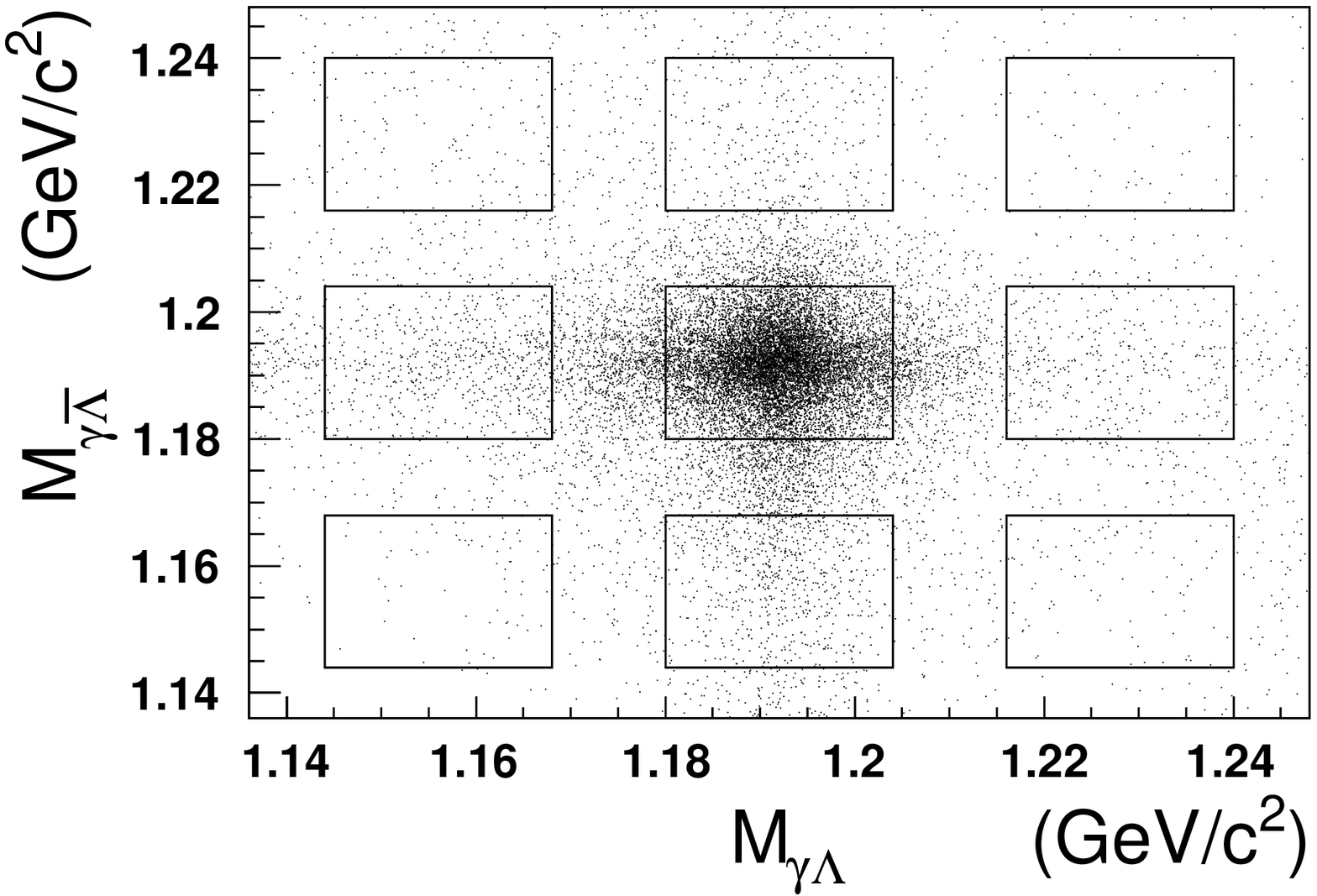}
\end{minipage}}%
\caption{The scatter plot of $M_{\gamma\Lambda}$ versus
$M_{\gamma\lamb}$ for $\jpsi\ar\ssb$ for (left) data and (right) inclusive
MC events.}
\label{fig7}
\end{figure}

\item Efficiency of the kinematic fit. The control sample of
$\jpsi\ar\Sigma^0\bar{\Sigma^0},\Sigma^0~(\bar{\Sigma}^0)\ar\gamma\Lambda~(\lamb)$
is used to study the efficiency of the 4C kinematic fit since its final
state is the same as our signal.
The event selection criteria for charged tracks and photons and the
reconstruction of $\Lambda~(\lamb)$ are the same as in our
analysis. If there are more than two photon candidates in an event, we loop
over all possible combinations and keep the one with the smallest value
for $(M_{\gamma\Lambda}-M_{\Sigma^0})^2
+(M_{\gamma\lamb}-M_{\bar{\Sigma}^0})^2$. Furthermore, the remaining
backgrounds are suppressed by limiting the momentum windows of
$\Sigma^0$ and $\bar{\Sigma}^0$, i.e., $|P_{\Sigma^0}-980|<40$~MeV/$c$
and $|P_{\bar{\Sigma}^0}-980|<40$ MeV/$c$. Figure~\ref{fig7} shows the
scatter plot of $M_{\gamma\Lambda}$ versus $M_{\gamma\lamb}$ for the
inclusive MC events and $\jpsi$ data after applying all event
selection criteria. The square in the center with a width of 10
MeV/$c^2$ is taken as the signal region. Almost no background is found
according to the topology analysis from inclusive MC events. The
candidate signal events for both data and MC events are subjected to
the same 4C kinematic fit as that in our analysis. The efficiency of
the 4C kinematic fit is defined as the ratio of the number of signal
events with and without a 4C kinematic fit. A correction factor,
$f_{4C}$, can be obtained by comparing the efficiency of the 4C kinematic
fit between data and MC simulation. i.e.,
$f_{4C}=\frac{\epsilon_{4C}^{data}}{\epsilon_{4C}^{MC}}$. The
efficiency corrections corresponding to $\chi^{2}<15,40,70$ are
$(90.3\pm0.8)$\%, $(97.5\pm0.6)$\% and $(98.7\pm0.3)$\%, respectively.
The errors in the efficiency corrections are taken as a systematic
uncertainty.

\item Fit range. The $\pi^0$, $\eta$, and $\Sigma(1385)^0$ yields are obtained by
fitting the data around the corresponding mass value. By changing the mass ranges for the fits,
the number of signal events changes slightly. These differences
are taken as the errors due to the uncertainty of the fit range.

\item Signal and sideband regions. By changing the signal and sideband region from 5 MeV/$c^2$$\times
5$ MeV/$c^2$ to 6 MeV/$c^2$$\times 6$ MeV/$c^2$, the number of fitted
$\pi^0,\eta$ and $\Sigma(1385)^0$ events changes slightly for data and MC.
The differences in yield between the two region sizes are taken as systematic errors.

\item Background shape. A part of the background depicted in Fig.~\ref{fit} is estimated by a fit
with a third-order Chebychev polynomial. The differences in signal yield with a background function that is
changed to a second-order polynomial, are taken as a systematic error due to the uncertainty in the description of the
background shape.

\item Total number of $\jpsi$ and $\psp$ events. The total numbers of $\jpsi$ and $\psp$ events are obtained from
inclusive hadronic $\jpsi$ and $\psp$ decays with uncertainties of
1.24\%~\cite{jpsinumber} and 0.81\%~\cite{psipnumber}, respectively.
\eitm
All the sources of systematic errors are summarized in Table~\ref{totjpsi}. The total
systematic error is calculated as the quadratic sum of all
individual terms.
\
\btbl[h]
\caption{Systematic errors in the measurements of the branching fractions (\%).}
\bcl
\doublerulesep 2pt
\small{
\begin{tabular}{cccccc}\hline\hline
Source&$\jpsi\ar\llbpi$&$\jpsi\ar\llbeta$&$\jpsi\ar\Sigma(1385)^0\lamb$+c.c.&$\psp\ar\llbpi$&$\psp\ar\llbeta$\\\hline\hline
Photon efficiency&2.0&2.0&2.0&2.0&2.0\\
Tracking efficiency &6.0&9.0&4.0&5.0&6.0\\
Vertex fit &2.0&2.0&2.0&2.0&2.0\\
Correction factor of 4C fit &0.6&0.3&0.6&0.8&0.6\\
Background function&0.6&0.2&1.5&negligible&2.5\\
Signal and sidebands&3.6&1.7&negligible&9.1&2.0\\
Fit range&0.6&0.4&negligible&negligible&1.5\\
${\cal B}(\Lambda\ar\pi p)$&0.8&0.8&0.8&0.8&0.8\\
${\cal B}(P\ar\GG)$&negligible&0.6&negligible&negligible&0.6\\
${\cal B}(\Sigma(1385)^0\ar\Lambda\pi^0)$&-&-&1.7&-&-\\
$N_{\jpsi}$&1.24&1.24&1.24&-&-\\
$N_{\psp}$&-&-&-&0.81&0.81\\\hline\hline
Total&7.8&9.7&5.6&10.9&7.6\\\hline\hline
\end{tabular}}
\label{totjpsi}
\ecl
\etbl

\section{Results}
The branching fraction of $\jpsi~(\psp)\ar X$ is determined by the relation
\[
{\cal B}(\jpsi~(\psp)\ar X)=\frac{N^{obs}[\jpsi~(\psp)\ar X\ar Y]}{N_{\jpsi~(\psp)}\cdot{\cal B}~(X\ar Y) \cdot\epsilon [\jpsi~(\psp)\ar X\ar Y]\cdot f_{4C}},
\]
and if the signal is not significant, the corresponding upper limit of
the branching fraction is obtained by
\[
{\cal B}(\jpsi~(\psp)\ar X)<\frac{N^{obs}_{UL}[\jpsi~(\psp)\ar X\ar Y]}{N_{\jpsi~(\psp)}\cdot{\cal B}~(X\ar Y)\cdot\epsilon [\jpsi~(\psp)\ar X\ar Y ]\cdot f_{4C}\cdot(1.0-\sigma_{sys.})},
\]
where, $N^{obs}$ is the number of observed signal events or its upper
limit $N^{obs}_{UL}$, $Y$ is the final state, $X$ is the intermediate
state, $\epsilon$ is the detection efficiency, and $\sigma_{sys.}$ is
the systematic error. The branching fraction of $X\ar Y$ is taken from
the PDG~\cite{PDG}. Table~\ref{numjpsi} lists the various numbers that were used in the
calculation of the branching fractions.
\begin{table*}[ht]
\caption{Numbers used in the calculations of the branching fractions.}
\bcl
\footnotesize{ \doublerulesep 2pt
\begin{tabular}{cccccc}\hline\hline
Channel&$\jpsi\ar\llbpi$&$\jpsi\ar\llbeta$&$\jpsi\ar\Sigma(1385)^0\lamb$+c.c.&$\psp\ar\llbpi$&$\psp\ar\llbeta$\\\hline\hline
Number of events $N^{obs}$/$N^{obs}_{UL}$&323&454&$<37$&$<9$&60.4 \\
Efficiency $\epsilon$~(\%)&9.65&8.10& 6.22&8.95&14.64\\
$f_{4C}~(\%)$ &97.5&98.7&97.5&90.3&97.5\\
$N_{\jpsi}~(\times10^6)$&225.3&225.3&225.3&-&-\\
$N_{\psp}~(\times10^6)$&-&-&-&106.41&106.41\\
${\cal B}(\Lambda\ar\pi p)~(\%)$&63.9&63.9&63.9&63.9&63.9\\
${\cal B}(\Sigma(1385)^0\ar\Lambda \pi^0)$~(\%)&-&-&87.5&-&-\\
${\cal B}(\pi^0,\eta\ar\GG)~(\%)$&98.8&39.4&98.8&98.8&39.4\\\hline\hline
\end{tabular}}
\label{numjpsi}
\ecl
\end{table*}
 With these, we obtain
\[
{\cal B}(\jpsi\ar\llbpi)= (3.78\pm0.27~{\rm (stat.)}\pm 0.29~{\rm (sys.)})\times
10^{-5},
\]
\[
{\cal B}(\psp\ar\llbpi)< 0.29\times 10^{-5},
\]
\[
{\cal B}(\jpsi\ar\llbeta)= (15.7\pm0.79~{\rm (stat.)}\pm 1.52~{\rm (sys.)})\times
10^{-5},
\]
\[
{\cal B}(\psp\ar\llbeta)= (2.47\pm 0.34~{\rm (stat.)}\pm
0.19~{\rm (sys.)})\times10^{-5},
\]
\[
{\cal B}(\jpsi\ar\Sigma(1385)^0\lamb+c.c)< 0.81\times 10^{-5}.
\]
Here the upper limits correspond to the 90\% C.L..

With these results, one can test whether the branching ratio between the
$\psp$ and $\jpsi$ decays to the same hadronic final state, $Q_h$,
is compatible with the expected 12\% rule~\cite{rule}. We find a
$Q_{h}$ for the channels $\llbpi$ and $\llbeta$ of
\[
Q_h=\frac{{\cal B}(\psp\ar\llbpi)}{{\cal
B}(\jpsi\ar\llbpi)}<10.0\%~\] at the 90\%~C.L., and,
\[
Q_h=\frac{{\cal B}(\psp\ar\llbeta)}{{\cal
B}(\jpsi\ar\llbeta)}=(15.7\pm 2.9)\%.
\]
The errors reflect a quadratic sum of the systematic and statistical error, whereby
some of the common sources of systematic errors have been canceled. Clearly, the isospin-violated decay $\llbpi$ is
suppressed in $\psp$ decays, while $Q_h$ for the isospin-allowed decay, $\llbeta$, agrees with the \textquotedblleft 12\%" rule within about 1$\sigma$.

\section{Summary}
This paper presents measurements of the branching fractions of the
isospin-violating and isospin-conserving decays of the $\jpsi$ and
$\psp$ into $\llbpi$ and $\llbeta$, respectively. The results
together with the measurements from previous experiments are
summarized in Table~\ref{sum}. We note that the earlier measurements
of the branching fraction of the decay $\llbpi$ by BESI and DM2
likely overlooked a sizeable background contribution in their
analysis as supported by the BESII and BESIII results. Hence, we
claim that we have observed for the first time the two processes,
$\jpsi\ar\llbpi$ and $\psp\ar\llbeta$. Moreover, the branching
fractions of the $\jpsi\ar\llbeta$ decay is measured with a
drastically improved precision. Its central value is lower than the
BESII measurement by about 1.5$\sigma$. The branching ratios of
$\jpsi\ar\llbpi$ and $\psp\ar\llbeta$ are consistent with previous
upper limits, and the upper limit of $\psp\ar\llbpi$ is
significantly more stringent than the BESII measurement. The
isospin-violating decay modes, $\jpsi\ar\llbpi$ and $\psp\ar\llbpi$,
are suppressed relative to the corresponding isospin-conserving
decay modes into $\llbeta$, albeit only by a factor of 4 in the case
of the $\jpsi$ decay. In addition, we search for the isospin
violating decays of $\jpsi\ar\Sigma(1385)^0\lamb+c.c.$ and no
significant signal is observed.


\btbl[h]
\caption{\footnotesize{A comparison of the branching fractions of this work with the results of previous experiments~($\times 10^{-5}$). The first error is statistical and the second one indicates the systematical uncertainty.}}
\bcl
\footnotesize{
\doublerulesep 2pt
\begin{tabular}{lcccc}\hline\hline
Experiments&${\cal B}(\jpsi\ar\llbpi)$&${\cal B}(\jpsi\ar\llbeta)$&${\cal B}(\psp\ar\llbpi)$&${\cal B}(\psp\ar\llbeta)$\\\hline\hline
This experiment&$3.78\pm 0.27\pm 0.29$&$15.7\pm 0.79\pm 1.52$&$<0.29$&$2.47\pm 0.34\pm 0.19$\\
BESII~\cite{xuxp}&$<6.4$&$26.2\pm 6.0\pm4.4$&$<4.9$&$<12$\\
BESI~\cite{bes1}&$23.0\pm 7.0\pm 8.0$&&&\\
DM2~\cite{dm2}&$22.0\pm 5.0\pm 5.0$&&&\\\hline\hline
\end{tabular}}
\label{sum}
\ecl
\etbl
\section{acknowledgment}
The BESIII collaboration thanks the staff of BEPCII and the computing center for their hard efforts. This work is supported in part by the Ministry of Science and Technology of China under
Contract No. 2009CB825200; National Natural Science Foundation of
China (NSFC) under Contracts Nos. 10625524, 10821063, 10825524,
10835001, 10935007, 11125525, 10975143, 11079027, 11079023; Joint
Funds of the National Natural Science Foundation of China under
Contracts Nos. 11079008, 11179007; the Chinese Academy of Sciences
(CAS) Large-Scale Scientific Facility Program; CAS under Contracts
Nos. KJCX2-YW-N29, KJCX2-YW-N45; 100 Talents Program of CAS;
Istituto Nazionale di Fisica Nucleare, Italy; Ministry of
Development of Turkey under Contract No. DPT2006K-120470; U. S.
Department of Energy under Contracts Nos. DE-FG02-04ER41291,
DE-FG02-91ER40682, DE-FG02-94ER40823; U.S. National Science
Foundation; University of Groningen (RuG) and the Helmholtzzentrum
fuer Schwerionenforschung GmbH (GSI), Darmstadt; WCU Program of
National Research Foundation of Korea under Contract No.
R32-2008-000-10155-0.

\end{document}